 \documentclass{jfm}

\makeatletter
 \gdef\@underjournal{}
\def\@maketitle#1{%
 \newpage
 \vspace*{10\p@}%
 {\centering \sloppy
  {\normalfont\LARGE\fontswitch\bfseries \@title \par}%
  \vskip 14\p@ \@plus 2\p@ \@minus 1\p@
  {\normalfont\large\fontswitch\bfseries\baselineskip=12\p@
     \lowercase{\@author}\par}%
  \vskip 4\p@ \@plus 1\p@
  {\normalfont\small \@affiliation \par}%
  \vskip 8\p@ \@plus 2\p@ \@minus 1\p@
 \par}%
 \vskip 8\p@ \@plus 2\p@ \@minus 1\p@
}
\makeatother

\usepackage[svgnames,table]{xcolor}
\usepackage{amsmath,amssymb}
\usepackage{mathrsfs}
\usepackage{array}
\usepackage{graphicx}
\usepackage{natbib}

\providecommand\bnabla{\boldsymbol{\nabla}}
\newcommand{\bvort}{\boldsymbol{\omega}}

\newcommand\Rey{\mbox{\textit{Re}}}
\newcommand\Ro{\mbox{\textit{Ro}}}
\newcommand\Fr{\mbox{\textit{Fr}}}
\newcommand\Bu{\mbox{\textit{Bu}}}
\newcommand\Pran{\mbox{\textit{Pr}}}
\newcommand\BV{Brunt-V\"ais\"al\"a frequency}

\renewcommand{\vec}[1]{\mathbf{#1}}
\newcommand{\wavevecset}[1]{\mathcal{#1}}
\newcommand{\fourier}[1]{\hat{#1}}
\usepackage{mathtools}
\DeclarePairedDelimiter\abs{\lvert}{\rvert}

\title[Waves, vortices, and the inverse cascade of rotating-stratified turbulence]{Waves and vortices in the inverse cascade regime of stratified turbulence with or without rotation}

 \author[C.~Herbert, R.~Marino, D.~Rosenberg and A.~Pouquet]{Corentin Herbert$^{1,2}$, Raffaele Marino$^{2,3,4}$, Duane Rosenberg$^5$ and Annick Pouquet$^{2,6}$}

\affiliation{
$^{1}$Department of Physics of Complex Systems, Weizmann Institute of Science, 76100 Rehovot, Israel.\\
$^{2}$National Center for Atmospheric Research, P.O.~Box 3000, Boulder, CO 80307, USA.\\
$^{3}$Space Sciences Laboratory, University of California, 7 Gauss way, Berkeley, CA 94720, USA. \\
$^{4}$Dipartimento di Fisica, Universit\`a della Calabria, Ponte P. Bucci, cubo 31C, 87036 Rende, Italy.\\
$^{5}$National Center for Computational Sciences, Oak Ridge National Laboratory, P.O. Box 2008, Oak Ridge, TN 37831, USA. \\
$^{6}$Laboratory for Atmospheric and Space Physics, University of Colorado, Boulder, CO 80309, USA.}
\date{\today}

\begin{document}

\maketitle

\begin{abstract}
We study the partition of energy between waves and vortices in stratified turbulence, with or without rotation, for a variety of parameters, focusing on the behavior of the waves and vortices in the inverse cascade of energy towards the large scales.
To this end, we use direct numerical simulations in a cubic box at a Reynolds number $\Rey \approx 1000$,
 with the ratio between the \BV~$N$ and the inertial frequency $f$ varying from 1/4 to 20, together with a purely stratified run.
The Froude number, measuring the strength of the stratification, varies within the range $0.02 \le \Fr \le 0.32$.
We find that the inverse cascade is dominated by the slow quasi-geostrophic modes. 
Their energy spectra and fluxes exhibit characteristics of an inverse cascade, even though their energy is not conserved.
Surprisingly, the slow vortices still dominate when the ratio $N/f$ increases, also in the stratified case, although less and less so.
However, when $N/f$ increases, the inverse cascade of the slow modes becomes weaker and weaker, and it vanishes in the purely stratified case.
We discuss how the disappearance of the inverse cascade of energy with increasing $N/f$ can be interpreted in terms of the waves and vortices, and identify three major effects that can explain this transition based on inviscid invariants arguments.
\end{abstract}

\section{Introduction}

The interaction between vortices and waves is a long-standing problem in fluid turbulence. 
It can lead to a self-sustaining process that is dominant, for example in pipe flows, due to the strong wall-induced shear. 
A quasi-linear analysis around a mean but unsteady flow allows for a prediction of large-scale coherent structures in such flows~\citep{McKeon2010, Farrell2012, Sharma2013}, to the prediction of jets in baroclinic turbulence in the atmosphere of planets~\citep{Farrell2009}, and it can also be used as a tool to control the onset of turbulence using traveling waves~\citep{Moarref2010}.
The dynamics of geophysical flows involves complex interactions between nonlinear eddies and waves due to the combination of planetary rotation and density stratification as well as the formation of shear layers. 
Typical parameters for the atmosphere and the ocean at large scales (larger than about 1000 km) are such that the dynamics of the fast waves is slaved to that of the eddies, and this leads to a quasi-geostrophic quasi-linear regime characterized by a balance between pressure gradient and the Coriolis and gravity forces~\citep{Charney1971}. 
\cite{Augier2013} have shown using general circulation models (GCMs) that the ratio between the energy in the balanced vortices and the energy in the horizontally divergent motions (including internal waves) was on the order of 100.
Then, the eddy turn-over time is much larger than typical wave frequencies: the \BV~$N$ or the inertial frequency $f=2\Omega$ (with $\Omega$ the rotation rate). 
For instance, frequency spectra obtained from in-situ measurements in the ocean show distinct power-laws for internal waves in the $[f,N]$ band and geostrophic eddies at lower frequencies~\citep{Ferrari2009}. 
The quasi-geostrophic regime is well understood theoretically and provides a good description of the large scales of the atmosphere and the ocean~\citep[e.g.]{VallisBook}.
However, the range of scales in these geophysical flows before dissipation prevails at small scales is such that other scale-dependent regimes can arise in which turbulence comes into play, with the eddy turn-over time becoming comparable to the \BV, the inertial frequency, or their combinations which arise in the dispersion relation of internal waves. 
\cite{Lilly1983}, in particular, has underlined the emergence of a regime, coined \emph{stratified turbulence}, between quasi-geostrophic turbulence and homogeneous isotropic turbulence (see also the scale analysis by~\cite{Riley1981}), and its importance for the mesoscales (with horizontal scales between 1 and 100 km, approximately) in the atmosphere. 
Indeed, the measured horizontal spectra of kinetic and potential energy in the upper troposphere and lower stratosphere~\citep{Nastrom1984} has a long history of opposite interpretations: some have claimed that this unambiguous $k^{-5/3}$ range corresponds to an inverse cascade of quasi-2D eddies~\citep{Gage1979,Lilly1983,Metais1996}, while others have argued that it was in fact a direct energy cascade dominated by internal waves~\citep{Dewan1979}.
This example emphasizes the importance of analyzing data --- be it observational, experimental or numerical --- in terms of eddies or vortices on the one hand, and waves on the other hand. 
In this paper, we shall try to disentangle these two types of motion in the idealized context of direct numerical simulations of stratified flows with and without rotation in a periodic box, focusing on the classical quantities characterizing the statistical behavior of nonlinear interactions in turbulent flows, in particular energy spectra and transfer functions.

One way to distinguish between waves and eddies is through a decomposition of the flow onto the normal modes of the linearized system~\citep[see also below, \S \ref{S:NORMAL}]{Bartello1995}. 
Normal mode decomposition is a standard technique which has been used in a variety of theoretical~\citep{Errico1984,Warn1986,Lien1992,Bartello1995,Waite2004,Herbert2014c} and numerical studies~\citep{Bartello1995,Waite2006b,Waite2006a,Sukhatme2008,Kitamura2010,Kurien2012,Kurien2014,Brunner2014}, for the Boussinesq equations, which we shall investigate here, as well as for other dynamical equations.
Recently, another approach has been proposed~\citep{Buhler2014}, which relies on a few additional assumptions in order to be able to achieve a similar decomposition without a complete knowledge of the dynamical fields. 
This is particularly useful when working with observational data, such as ship track data or airplane measurements. \cite{Callies2014} have used this method to claim that observational data supported the internal wave interpretation of the forward cascade of the atmospheric mesoscales, although~\cite{Lindborg2015} argues using a similar technique that the forward cascade of energy could in fact be of another type, corresponding to the scaling regime of stratified turbulence identified in~\cite{Billant2001} (see also \citep{Lindborg2006}).
Another interesting approach consists of computing a full frequency-wavenumber energy spectrum in the four dimensional Fourier space (\cite{ClarkdiLeoni2015}; see also~\cite{Campagne2015} for an analogous analysis in a laboratory experiment of rotating flow). 
This allows to directly assess how the energy is partitioned between modes which exhibit phase coherence, like waves, and other modes. 
However, this is a computationally expensive technique which requires a good time resolution in addition to a good space resolution (and hence, a lot of disk space).

In this context, a pioneering analytical and numerical study of strongly rotating stratified turbulence in terms of normal modes was done by~\cite{Bartello1995}.
The link between geostrophic adjustment at large scale and the inverse cascade of energy due to the conservation of (linearized) potential vorticity, is clearly established through a decoupling of the (fast) wave modes and vortical (slow) modes. 
By truncating the nonlinear interactions to some specific subsets of modes, as for example between three slow modes that lead to the inverse cascade of energy~\citep{Charney1971}, dominant effects can thus be identified. It is shown in particular that there are nonlinear interactions using a slow mode as a catalyst to push the wave energy to small scales (whereas  triads involving three fast modes are negligible); furthermore, as the rotation is increased, the energetic exchanges between the slow vortical modes and the fast wave modes are weakened and such modes can be seen as actually being progressively decoupled. 
It was noticed by~\cite{LSmith2002} that, for long times, the growth of large-scale energy is associated with geostrophic modes for $1/2\le N/f \le 2$ (in which domain three-wave resonances vanish completely), whereas for large $N/f$, the ageostrophic thermal winds, also called vertically sheared horizontal flows (VSHF), are  dominant, with no variation in the horizontal and no vertical velocity.
One should note  that if the forcing wavenumber is not well separated from the box overall dimension, finite-size effects can develop that quench the possible (discrete) resonances, as already noted in the context of water-wave  turbulence (see e.g.~\cite{Kartashova2008} and references therein).

In the purely stratified case, it was found by~\cite{Waite2004, Waite2006a} that the vertical spectrum of the waves moves to larger vertical wavenumber as stratification increases, in agreement with the argument found by~\cite{Billant2001} that the characteristic scale in the vertical is proportional to $U/N$, with $U$ a typical large-scale velocity;  at smaller scales the structures break-down in a nonlinear fashion although isotropy 
may not be restored, depending on the buoyancy Reynolds number and whether the Ozmidov scale is resolved (see \S~\ref{ss:scales} below for definitions):
 the buoyancy scale $L_B=U/N$ must be sufficiently resolved for over-turning to take place.
The case of comparable (but not identical) strengths for rotation and stratification was analyzed by~\cite{Sukhatme2008}, where it is shown that there is a transition in behavior for $N/f\approx 1$: for smaller values, large scales are dominated by the vortical mode and small scales by wave modes,
whereas for larger values of $N/f$, the wave modes dominate at all scales smaller than the forcing.

The  scaling behavior of slow and fast modes for rotating stratified flows is analyzed in~\cite{Kurien2012, Kurien2014}  for regimes where 
potential vorticity is linear, for small aspect ratio of the domain size in the vertical ($\delta=L_\parallel/L_\perp =1/4$), but with a unit Burger number $Bu=\delta N/f$, using high resolution direct numerical simulations and hyper-viscosity. 
When compared to similar flows with $\delta=1$, differences arise, as for example when examining the steepness of the wave component of the energy spectrum, viewed as a signature of the layered structure of the flow.
The case $\delta\not= 1$ was also examined for either purely rotating~\citep{Deusebio2014a} or purely stratified flows~\citep{Sozza2015}, where it is shown that the aspect ratio can affect the strength of the inverse cascade of energy to large scales.
The combined effect of varying $N/f$ and $\delta$ on the formation of characteristic structures was further studied in~\cite{Kurien2014} for low Froude number ($Fr=0.002$, ensuring that potential vorticity is again linear, i.e. for low effective turbulence) and varying either the Rossby number at fixed $\delta$ or vice-versa (in which case, the choice was made of N=f). 
The layering of the flow is clearly dependent on the strength of the rotation ($Fr$ being fixed), and the wave and slow-mode spectra are quite different (when the total spectra are more difficult to distinguish).
The inverse cascade of energy is clearly associated with the vortical (slow) mode, as shown as well in~\cite{Brunner2014} for a variety of forcing functions in the context of sub-mesoscale oceanic mixing. 
It is also shown in~\cite{Brunner2014} that small-scale (shear) layers do not directly modify the inverse cascade although it could strengthen it when bringing together nearby vortices through nonlinear advection when they are sufficiently numerous.

Our study uses similar methods as the above mentioned authors, but pursues a slightly different goal: here, we focus on the possibility of an inverse cascade and its relation with the inviscid invariants of the system. 
We are interested in understanding the difference between the rotating-stratified case, for which an inverse cascade is possible~\citep{Marino2013b}, and the purely stratified case, where there is no such thing (although there is some transfer of energy towards the large scale). 
This difference has already been considered from the point of view of anisotropic energy transfers~\citep{Marino2014}; here we adopt a different point of view and we investigate the role of vortices and waves in the inverse cascade (or lack thereof).
We also aim at characterizing the transition between these two regimes. Indeed, although previous studies have established that when an inverse cascade is present, it is due to the slow vortices, it is not clear how this phenomenology breaks up. 
We shall see that in the regime of parameters considered, we can identify three potential candidates, which can also be superimposed.
To do so, we examine data stemming from direct numerical simulations of rotating stratified turbulence for a variety of $N/f$ ratios (and Froude and Rossby numbers) in cubic boxes at a Reynolds number of 1000. 
After describing the relevant equations and the numerical simulations in \S~\ref{numericssection}, we review the normal mode decomposition in \S~\ref{S:NORMAL}, and analyze the numerical data in terms of the normal modes in \S~\ref{S:TEMP}. In \S~\ref{discusssection}, we discuss the results, first in the framework of the inviscid invariants of the system, and then by considering the effect of the buoyancy Reynolds number on the energy partition between slow and fast modes. 
Finally, our conclusions are exposed in \S~\ref{S:CONCL}.

\section{Numerical Data}\label{numericssection}

\subsection{Equations of motion and direct numerical simulations}

We consider incompressible flows in a reference frame rotating with angular velocity $\Omega$ around axis $\vec{e}_z$ (we note $\vec{\Omega}=\Omega \vec{e}_z$ and $f=2\Omega$ the Coriolis parameter). 
The gravity field is anti-aligned with the direction of rotation: $\vec{g}=-g\vec{e}_z$. 
We work in the Boussinesq approximation: the density is assumed constant, equal to $\rho_0>0$, except in the buoyancy force term, and we assume a linear background stratification: the density field is given by $\rho(\vec{x})=\rho_0\left(1-\frac{N^2}{g}z\right)+\rho'(\vec{x})$, with $N$ the Brunt-V\"ais\"al\"a frequency.
Note that we are only considering the case of stably stratified flows.
We introduce the \emph{temperature} field (with the dimension of a velocity) $\theta(\vec{x})=\frac{g}{N\rho_0}\rho'(\vec{x})$.
The velocity field is denoted by $\vec{u}$. 
Given the assumptions mentioned above, the equations governing the dynamics of the fields $\vec{u},\theta$ are as follows:
\begin{align}
\partial_t \vec{u} +\vec{u}\cdot \nabla \vec{u} &= - \nabla P + \nu \Delta \vec{u} - 2\Omega \vec{e}_z \times \vec{u} -N\theta \vec{e}_z + \vec{F},\label{bousseq1}\\
\partial_t \theta + \vec{u} \cdot \nabla \theta &= N u_z + \kappa \Delta \theta,\label{bousseq2}\\
\nabla \cdot \vec{u} &=0,\label{bousseq3}
\end{align}
where $P$ denotes the (rescaled) pressure field, fixed by the incompressibility condition, $\nu$ the viscosity and $\kappa$ the thermal diffusivity; we work with Prandtl number equal to unity: $\Pran=\nu/\kappa=1$. 
Finally, $\vec{F}$ is an isotropic forcing introduced only in the momentum equation; it has a fixed amplitude in a prescribed narrow shell centered on wavenumber $k_F$, and random but constant in time phases. 
Note that, contrary to other studies~\citep[for instance]{Kurien2012,Kurien2014}, there is no forcing in the buoyancy equation, but the forcing has a vertical component (unlike, e.g.~\cite{Waite2004} or~\cite{Sozza2015}) and does not satisfy balance relations, even at the lowest order (unlike e.g.~\cite{Waite2006a} where the forcing is only in the slow modes). 
The projection of the forcing on the waves and slow modes (see \S~\ref{S:NORMAL}) has the same magnitude; hence we are not favoring any of these modes \emph{a priori}.

We integrate equations (\ref{bousseq1}-\ref{bousseq3}) numerically with a pseudo-spectral code, the Geophysical High-Order Suite for Turbulence (GHOST) in a tri-periodic cubic domain of length $L=2\pi$. {GHOST} is parallelized using an hybrid  MPI/OpenMP method and scales linearly up to 10$^5$ compute cores on grids of up to 6144\textsuperscript{3} points~\citep{Mininni2011b}. 
No model is used for sub-grid scales, and dissipation occurs through standard Laplacian terms.
The runs considered here all have a resolution of 512\textsuperscript{3}, with forcing at wave number $k_F \in [22,23]$, and therefore Reynolds numbers $\Rey \approx 10^3$. 
The Rossby and Froude numbers vary as indicated in Table \ref{runstable}; the runs can be grouped in two series, one with constant Froude number ($\Fr=0.04$) and varying $N/f$ (and therefore $\Ro$) and one with constant Rossby number ($\Ro=0.08$) and varying $N/f$ (and therefore $\Fr$).
 The isotropic energy spectra and fluxes of the above runs have already been analyzed in~\cite{Marino2013b}, providing evidence for the existence of an inverse cascade of energy in the presence of rotation and stratification, but not in the purely stratified case (see also~\cite{Marino2014}).
 
 \begin{table}
 \centering
\begin{tabular}{c>{$}c<{$}>{$}c<{$}>{$}c<{$}>{$}c<{$}>{$}c<{$}>{$}c<{$}>{$}c<{$}>{$}c<{$}>{$}c<{$}>{$}c<{$}}
Id   & n_p & f & N & \Fr & \Ro & \Bu & \Rey & \mathcal{R}_B & R_\omega & k_F \\
\hline
R$1/4$   & 512^3 & 44 & 11 & 0.32 & 0.08 & 0.25 & 1000 & 102.4 & 2.53 & 22\\
\rowcolor{Tan!15} R$1/2$ & 512^3 & 44 & 22 & 0.16 & 0.08 & 0.5 & 1000 & 25.6 & 2.53 & 22\\
\rowcolor{Tan!15} R$3/2$ & 512^3 & 44 & 66 & 0.05 & 0.08 & 1.5 & 1000 & 2.5 & 2.53 & 22\\
R3  & 512^3 & 44 & 132 & 0.027 & 0.08 & 3 & 1000 & 0.73 & 2.53 & 22\\
R4 & 512^3 & 44 & 178 & 0.02 & 0.08 & 4 & 1000 & 0.4 & 2.53 & 22\\
\rowcolor{Tan!15} F$1/2$ & 512^3 & 178 & 89 & 0.04 & 0.02 & 0.5 & 1000 & 1.6 & 0.63 & 22\\
\rowcolor{Tan!15} F$3/2$ & 512^3 & 58 & 89 & 0.04 & 0.06 & 1.5 & 1000 & 1.6 & 1.90 & 22\\
\rowcolor{Tan!15} F2      & 512^3 & 44 & 89 & 0.04 & 0.08 & 2 & 1000   & 1.6 & 2.53 & 22\\
F3    & 512^3 & 29 & 89 & 0.04 & 0.12 & 3 & 1000  & 1.6 & 3.80& 22\\
F4   & 512^3 & 22 & 89 & 0.04 & 0.16 & 4 & 1000   & 1.6 & 5.06 & 22\\
F7   & 512^3 & 13 & 89 & 0.04 & 0.28 & 7 & 1000     & 1.6 & 8.85 & 22\\
F10   & 512^3 & 9 & 89 & 0.04 & 0.4 & 10 & 1000     & 1.6 & 12.65& 22\\
F20   & 512^3 & 4.5 & 89 & 0.04 & 0.8 & 20 & 1000  & 1.6 & 25.3& 22\\
F$\infty$   & 512^3 & 0 & 89 & 0.04 & \infty & \infty & 1000  & 1.6 & \infty & 22\\
\hline \end{tabular}
\caption{Model run parameters; names of runs at fixed Rossby number (\textit{resp.} fixed Froude number) begin with a R (\textit{resp.} F), followed by its N/f value.
$n_p$ is the number of grid points, N and f are the Brunt-V\"ais\"al\"a and the inertial frequency (with $f=2\Omega$, $\Omega$ being the imposed rotation), $Fr=U/[LN]$ and $Ro=U/[Lf]$ are the Froude and Rossby numbers based on large-scale data, and the Burger number is defined as $Bu=N/f$ for a cubic box as used here; $\mathcal{R}_B=\Rey \Fr^2$ and 
$R_\omega = \Rey^{1/2} \Ro =\omega_{rms}/f$ are, respectively, the buoyancy Reynolds number and the micro-Rossby number, and $k_F$ is the forcing wavenumber. 
The runs belonging to the no-resonance zone have a shaded background.}  
\label{runstable} 
\end{table}

\subsection{Characteristic scales} \label{ss:scales}
Rotating stratified turbulence covers a wide range of physical regimes, which can be characterized by three non-dimensional numbers, besides $Pr$. 
Given a characteristic length scale $L$ and a characteristic velocity $U$ at that scale, we introduce as usual the Reynolds number $\Rey=UL/\nu$, the Rossby number $\Ro=U/(fL)$ and the Froude number $\Fr=U/(NL)$. 
The Rossby and Froude numbers quantify the strength of the Coriolis and buoyancy forces, respectively. 
They correspond to the ratio between a wave period ($f^{-1}$ and $N^{-1}$, respectively) and the eddy turnover time $L/U$. 

A difficulty in rotating-stratified flows arises from the variety of characteristic scales one can construct. 
To start with, the Froude number can be based on a characteristic horizontal length scale or a vertical length scale. 
It is known that stratified turbulence develops strong gradients in the vertical (leading to the ubiquitous layered structure of the flow, see e.g.~\cite{Herring1989}), in such a way that the vertical Froude number remains of order unity~\citep{Billant2001,Lindborg2006}, even when the horizontal Froude number is small.
The corresponding vertical length scale is the buoyancy scale, $L_B=U/N$ and can be identified readily in such flows (see e.g., \cite{Rorai2014}). 
In the presence of rotation, the layers are expected to be slanted (with respect to the horizontal).
In that case, it is conjectured by~\cite{Billant2001} that the characteristic length scale in the vertical also involves both $N$ and $f$, like a deformation radius (for which the effects of rotation and stratification balance each other, with $Nk_\perp \sim f k_\parallel$). 
It was shown by \cite{Rosenberg2015} that a Froude number based on a vertical Taylor scale involving vertical velocity gradients, $\ell_z=\sqrt{\langle u_\perp^2 \rangle / \langle (\partial_z u_\perp)^2 \rangle}$ (where the angular brackets denote spatial averaging) is also very close to unity.
Here, we define the Froude and Rossby numbers (as well as the Reynolds number) based on the scale of the forcing, $L_F=2\pi/k_F$. 
Since the forcing is isotropic, there is no reason to distinguish between vertical and horizontal Froude number with this definition, although anisotropies and different characteristic scales in the horizontal and the vertical will of course develop spontaneously in the flow. 
It follows that for the purpose of our study, at unit aspect ratio, the ratio of the Rossby and Froude numbers $\Bu=\Ro/\Fr$, referred to as the Burger number, reduces to the ratio of the Brunt-V\"ais\"al\"a frequency and the Coriolis parameter: $\Bu=N/f$.

The recovery of isotropy should occur at a scale for which the eddy turn-over time and the wave period become comparable.
This allows one to define an Ozmidov  and a Zeman scale~\citep{Zeman1994,Mininni2012} written, respectively, for purely stratified or purely rotating flows, assuming an isotropic Kolmogorv range, $E_V(k)\sim \epsilon_V^{2/3}k^{-5/3}$ at small scale, as:
\begin{align}
\ell_{Oz}&= \sqrt{ \frac{\epsilon_V}{N^3} },& \ell_{Ze}&= \sqrt{ \frac{\epsilon_V}{f^3}},
\end{align}
with $\epsilon_V=\nu \langle \bvort^2 \rangle$ the kinetic energy dissipation rate.
The Froude and Rossby numbers measure the ratio of these scales to the characteristic scale: $\Fr=(\ell_{Oz}/L)^{2/3}, \Ro=(\ell_{Ze}/L)^{2/3}$.
Finally, if one wants to  construct a measure of the actual development of small-scales in the flow, a buoyancy Reynolds number $\mathcal{R}_B$ and a micro Rossby number $R_\omega$ can be defined in the following way: 
\begin{align}
\mathcal{R}_B &= \Rey \Fr^2,& R_\omega &= \frac{\omega_{rms}}{f} = Re^{1/2} Ro,
\end{align}
so that  $R_\omega^2$ is the equivalent, for rotation, of the buoyancy Reynolds number.

With these definitions, and recalling the Kolmogorov dissipation scale $\ell_\eta=(\nu^3/\epsilon_V)^{1/4}$, one can show that $\mathcal{R}_B=(\ell_{Oz}/\ell_\eta)^{4/3}$. 
In other words, when one wants to measure the intensity of stratified turbulence, the Ozmidov scale plays the role of the integral scale and $\mathcal{R}_B$ that of the Reynolds number.
In particular, when $\mathcal{R}_B$ is of order one, like in most of the runs shown here (see Table~\ref{runstable}), the Ozmidov scale coincides with the dissipation scale.
 Also note that, when $R_\omega \approx 1$, the small-scale turbulence is at least as vigorous as the imposed rotation frequency (see, e.g., \cite{SagautCambonBook}). These non-dimensional numbers are computed in Table~\ref{runstable}.

One of the goals of this paper is to study how the respective role of waves and vortices (see below) depends on the ratio $N/f$.
Note that there is a range that can be expected to have a special behavior: when $\frac 1 2 \leq \frac N f \leq 2$, there are no three-wave resonances (the simple proof relies on the fact that the inertia-gravity wave frequency $\sigma(\vec{k})$, defined by~\ref{dispersioneq}, is bounded by $f$ and $N$; see for instance~\cite{LSmith2002}, \S~6.1). 
Hence the no-resonance zone separates two regimes: $N/f < 1/2$ and $N/f > 2$. Here we shall focus on the latter range.

\subsection{Potential Vorticity}
\label{pvsection}

We define potential vorticity as usual:
\begin{align}
\Pi=  f\partial_z\theta -N\omega_z + \bvort \cdot \bnabla \theta,
\end{align}
where $\bvort=\bnabla \times \vec{u}$ is the standard vorticity.
Potential vorticity is a Lagrangian invariant of ideal Boussinesq flows: in the absence of forcing and dissipation, it is conserved along the trajectory of a fluid parcel, as expressed by the equation $\partial_t \Pi + \vec{u} \cdot \bnabla \Pi =0 $. 
Hence, it is analogous to vertical vorticity in 2D turbulence, a fundamental difference being that potential vorticity is not a linear functional of the dynamical fields, because of the quadratic term $\Pi_2=\bvort \cdot \bnabla \theta$, in addition to the linear term $\Pi_1=f\partial_z\theta -N\omega_z$. 
It follows that, unlike the 2D case, the $L_2$ norm of potential vorticity, referred to as potential enstrophy and denoted by $\Gamma$, is quartic and is not necessarily conserved by the Fourier truncation (see, e.g. \cite{Aluie2011}). 
\begin{figure*}
\centering
\includegraphics[width=\linewidth]{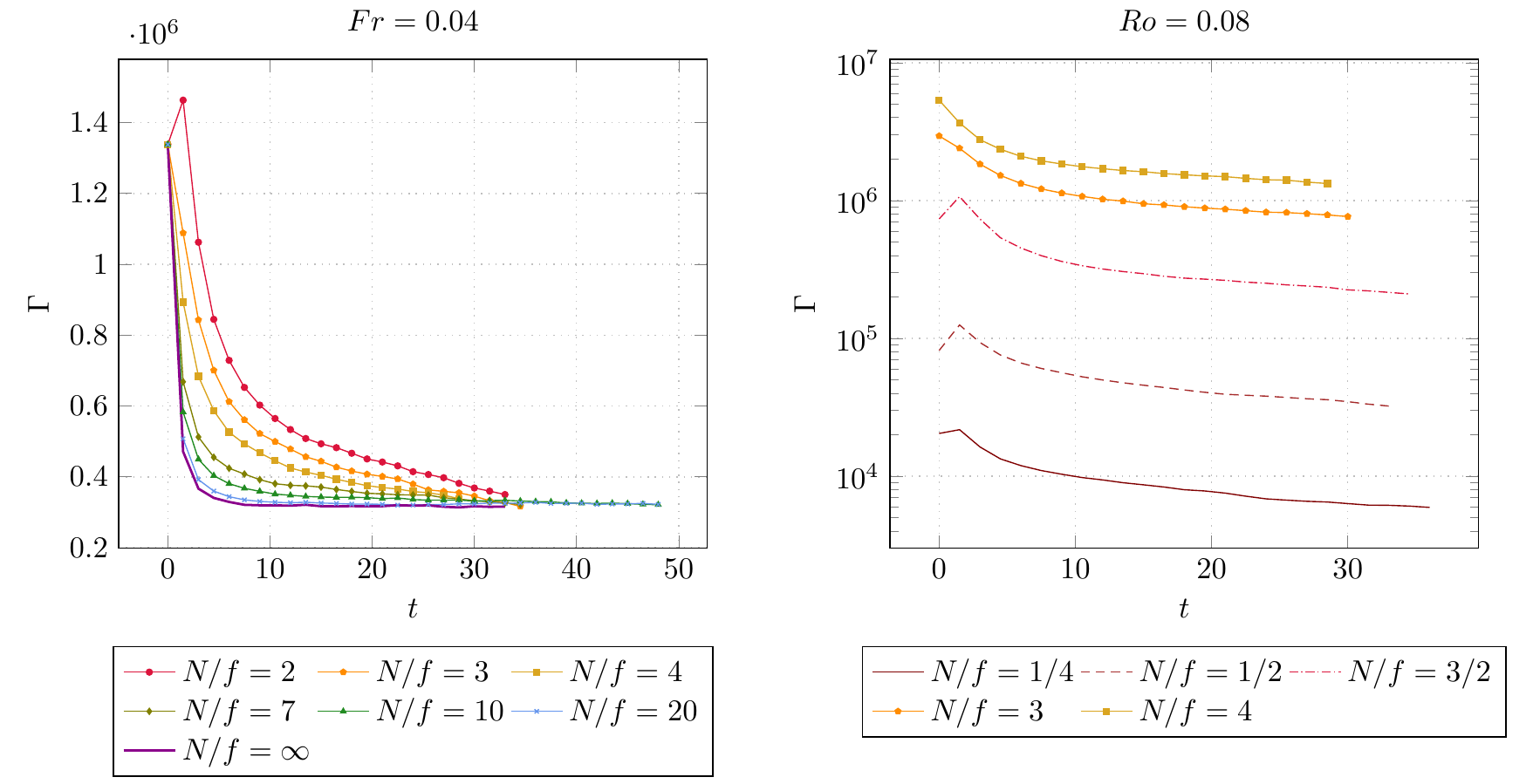}
\caption{Time evolution of the potential enstrophy for the series of runs at constant Froude number (left) and at constant Rossby number (right) with various $N/f$. The vertical scale is in logarithmic coordinates for the constant Rossby number series.}
\label{gammatotfig} 
\end{figure*}
We show in Fig.~\ref{gammatotfig} the time evolution of potential enstrophy in the two series of runs presented here. 
In the series of runs at constant Froude number, potential enstrophy decays to about 25\% of its initial value, then seems to remain constant in time. 
The decay is faster for higher $N/f$ ratios, but the final potential enstrophy level is about the same for all the runs. 
On the contrary, the series of runs at constant Rossby number exhibits very different levels of potential enstrophy, across about three orders of magnitude, and it increases with increasing $N/f$, i.e. decreasing Froude number. 
We conclude from these two graphs that, in the inverse cascade regime studied here, the value of potential enstrophy at late times is a function of the Froude number only. 
Also, potential enstrophy conservation is better satisfied when stratification is strong.
\begin{figure*}
\centering
\includegraphics[width=\linewidth]{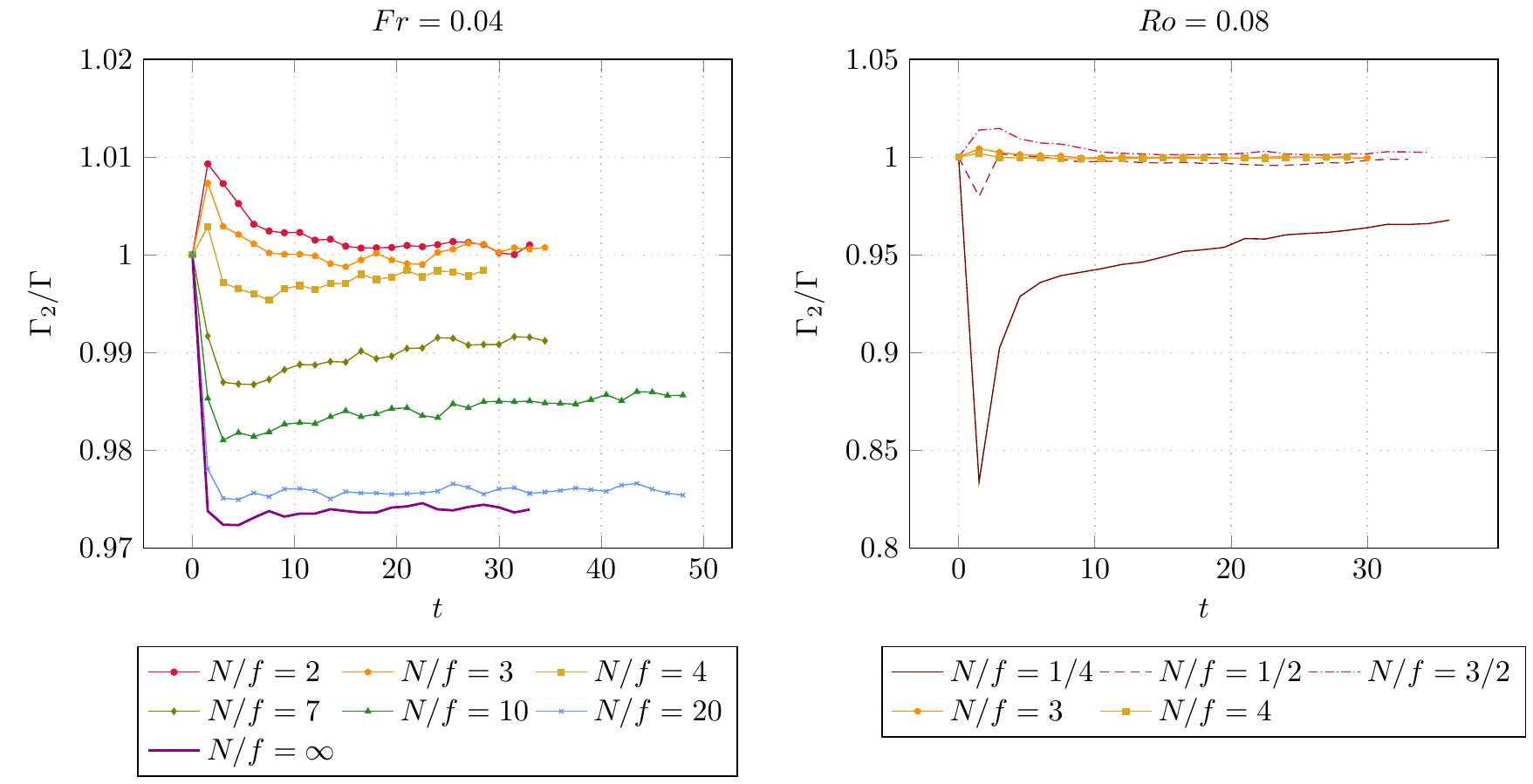}
\caption{Time evolution of the ratio of quadratic potential enstrophy to total potential enstrophy, for the series of runs at constant Froude number (left) and at constant Rossby number (right) with various $N/f$.}
\label{gammaratiofig}
\end{figure*}
We now introduce the quadratic, cubic and quartic components of potential enstrophy, defined respectively as
\begin{align}
\Gamma_2 &= \int \Pi_1^2,  \\
\Gamma_3 &= 2  \int  \Pi_1 \Pi_2,  \\
\Gamma_4 &= \int \Pi_2^2,
\end{align}
such that $\Gamma=\Gamma_2+\Gamma_3+\Gamma_4$. 
$\Gamma_{2},\Gamma_3$ and $\Gamma_4$ are respectively quadratic, cubic and quartic in the field variables, but they all have the same physical dimension, that of frequency square or square vorticity.
The ratio $\Gamma_2/\Gamma$ is shown in Fig.~\ref{gammaratiofig}. 
Note that this ratio can exceed unity, since the cubic component of total potential enstrophy is not positive definite (it measures correlations between the linear part of potential vorticity and the nonlinear part). 
This figure shows that in all the runs, total potential enstrophy is well approximated by its quadratic part, in agreement with the findings of~\cite{Kurien2014}, although it is less and less the case as rotation decreases (constant Froude series of runs) or in the $N/f=1/4$ (weak stratification) case, which also has the highest buoyancy Reynolds number $\approx 100$ (see Table \ref{runstable}).

In 2D turbulence, enstrophy conservation has very important consequences: it is ultimately responsible for the inverse cascade. 
It is a legitimate question to ask whether potential enstrophy conservation places such a strong constraint on rotating-stratified flows. 
Another important aim of this paper is to show that this constraint is strongly dependent on whether there is rotation or not.

\section{Waves and Vortices: Normal Mode Decomposition} \label{S:NORMAL}
\subsection{The linear dynamics}

 Inertia-gravity wave dynamics has been studied for a long time (see for example \cite{Hasselmann1966,Garrett1979,Muller1986,Bartello1995,Staquet2002}) and is summarized very briefly here in order to recall essential concepts and to define notations.
Indeed, the linear terms introduced by the Coriolis force and the buoyancy force in equations (\ref{bousseq1}-\ref{bousseq3}) allow for the propagation of dispersive waves. 
The linearized inviscid and unforced dynamics reads:
\begin{align}
\partial_t \vec{u}  &= - \nabla P  - f \vec{e}_z \times \vec{u} -N\theta \vec{e}_z,\label{linbousseq1}\\
\partial_t \theta  &= N u_z,\label{linbousseq2}\\
\nabla \cdot \vec{u} &=0.\label{linbousseq3}
\end{align}
This set of equations corresponds to the low Froude number limit ($\Fr \to 0$) of the Boussinesq equations, if we assume that the time scale is the buoyancy time scale $N^{-1}$~\citep{Lilly1983}.
The above equations admit travelling wave solutions of the form $u_j(\vec{x},t)=U_j e^{i(\vec{k}\cdot \vec{x}-\sigma(\vec{k}) t)}, \theta(\vec{x},t)=\Theta e^{i(\vec{k}\cdot \vec{x}-\sigma(\vec{k}) t)}, P= P_0 e^{i(\vec{k}\cdot \vec{x}-\sigma(\vec{k}) t)}$, for arbitrary wave vector $\vec{k}$, with relations between the complex amplitudes $U_j,\Theta,P_0$ fixed by the above equations, and with the dispersion relation 
\begin{align}
\sigma(\vec{k})=\sqrt{f^2 k_z^2/k^2+N^2 k_\perp^2/k^2}.\label{dispersioneq}
\end{align}
 Such waves are called \textit{inertia-gravity waves} (and include, of course, waves traveling in the $-\vec{k}$ direction). 
 Note that although the Coriolis parameter only appears in the evolution equation for the horizontal velocity, and the Brunt-V\"ais\"al\"a frequency only for the vertical velocity and temperature, inertia-gravity waves couple the four dynamical fields because of the incompressibility condition. 
 Without pressure, we would simply have inertial oscillations, at frequency $f$, for the horizontal velocity, decoupled from gravity waves of frequency $N$ acting on the vertical velocity and the temperature.

The nonlinear term introduces a coupling between waves with different wave vectors. 
A possible approach is to treat the dynamical fields as a wave field with slow nonlinear evolution of the magnitude of each wave mode, using a perturbative approach. 
This is a first kind of turbulence, referred to as {\it wave turbulence}~\citep{ZakharovBook,NazarenkoBook, Newell2011}.

Now, the set of linearized equations also admits stationary solutions, where the right hand side of equations (\ref{linbousseq1}-\ref{linbousseq2}) vanishes. 
These solutions have no vertical velocity ($u_z=0$), and the pressure gradient balances at the same time the Coriolis force ({\it geostrophic balance}, $\nabla_\perp P = - f \vec{e}_z \times \vec{u}$) and the buoyancy force ({\it hydrostatic balance}, $\partial_z P = -N\theta \vec{e}_z$). 
In other words, rotation and stratification act to maintain horizontal motion. 
This kind of balanced motion can also be seen as a low Froude number limit ($\Fr \to 0$) solution of the Boussinesq equations, assuming that $\Bu=1$ and that the time scale is the eddy turnover time~\citep{Lilly1983}. 
It corresponds to eddies of typical aspect ratio $f/N$, which are advected by the horizontal velocity field, much like vortices in 2D turbulence. 
Hence, another kind of turbulence, referred to as {\it geostrophic turbulence}, is encapsulated in the Boussinesq equations.

\subsection{The normal mode decomposition in Fourier space}

The two kinds of motion, waves and vortices, described above, correspond to normal modes of the linearized Boussinesq equations. 
Although they are coupled in the nonlinear dynamics, they provide a useful framework for analyzing the full system. 
Here, we shall try to disentangle the behavior of these two contributions in the simulated fields. 
To separate the two contributions, it is convenient to express them in Fourier space~\citep[see also \cite{Herbert2014c}]{Bartello1995}:
we introduce $\wavevecset{B}=\mathbb{Z}^3$ the set of wave vectors, and for each wave vector $\vec{k} \in \wavevecset{B}$, the vectors 
\begin{align}
\vec{Z}_0(\vec{k})&=\vec{M}(\vec{k}) 
\begin{pmatrix} 0 \\ 1 \\ 0\end{pmatrix},\quad \vec{Z}_-(\vec{k})=\vec{M}(\vec{k}) \begin{pmatrix} 1 \\ 0 \\ 0\end{pmatrix}, \quad\text{ and } \vec{Z}_+(\vec{k})=\vec{M}(\vec{k}) \begin{pmatrix} 0 \\ 0 \\ 1\end{pmatrix},
\end{align}
where the expression of the matrix $\vec{M}(\vec{k})$ is given by:
\begin{align}
\vec{M}(\vec{k})
&= 
\begin{cases}
\frac{1}{\sqrt{2} k k_\perp \sigma(\vec{k})}
\begin{pmatrix}
  f k_2 k_{\parallel}- i k_1 k_{\parallel} \sigma(\vec{k}) 	& - \sqrt{2} N k_2 k_{\perp} 		&  f k_2 k_{\parallel} + i k_1 k_{\parallel} \sigma(\vec{k}) \\
- f k_1 k_{\parallel}- i k_2 k_{\parallel} \sigma(\vec{k}) 		&  \sqrt{2} N k_1 k_{\perp} 		&  - f k_1 k_{\parallel}+ i k_2 k_{\parallel} \sigma(\vec{k}) \\
i k_{\perp}^2 \sigma(\vec{k})  						& 0 							& -i k_{\perp}^2 \sigma(\vec{k})  \\
 - N k_{\perp}^2 								& - \sqrt{2} f k_{\parallel} k_{\perp} 	& - N k_{\perp}^2 \\
\end{pmatrix}&
\text{if } k_\perp \neq 0,\\
\frac 1 {\sqrt{2}}
\begin{pmatrix}
i & 0 & -i\\
1 & 0 & 1\\
0 & 0 & 0\\
0 & -\sqrt{2} & 0
\end{pmatrix}&
\text{if } k_\perp=0.
\end{cases}
\end{align}
The vectors $\vec{Z}_0(\vec{k}),\vec{Z}_-(\vec{k})$ and $\vec{Z}_+(\vec{k})$ are mutually orthogonal, and normalized to unity:
\begin{align}
\vec{Z}_r(\vec{k})^\dagger \vec{Z}_s(\vec{k}) &= \delta_{rs},
\end{align}
where $\dagger$ denotes the transpose and complex conjugation. 
Although the matrix $\vec{M}(\vec{k})$ is rectangular, it satisfies the hermitian identity $\vec{M}(\vec{k})^\dagger \vec{M}(\vec{k})=I_3$.
They are the normal modes of the linearized dynamics: their evolution in the linearized dynamics is given by
\begin{align}
\vec{L}(\vec{k})\vec{Z}_0(\vec{k}) &= 0, \\
\vec{L}(\vec{k})\vec{Z}_+(\vec{k}) &= i\sigma(\vec{k})\vec{Z}_+(\vec{k}), \\
\vec{L}(\vec{k})\vec{Z}_-(\vec{k}) &= -i\sigma(\vec{k})\vec{Z}_-(\vec{k}),
\end{align}
where $\vec{L}(\vec{k})$ is the linear operator in Fourier space associated to the linearized dynamics give by Eqs. \ref{linbousseq1}-\ref{linbousseq3}.
We see that the mode $\vec{Z}_0(\vec{k})$ is conserved by the linear dynamics, while the modes $\vec{Z}_{\pm}(\vec{k})$, corresponding to the inertia-gravity waves, propagate with frequency $\sigma(\vec{k})$. 
Therefore, $\vec{Z}_0$, whose characteristic evolution time is the eddy turnover time, is referred to as the {\it slow mode}.
The normal modes can be used as a basis for the Fourier modes of the dynamical fields. 
Introducing the Fourier decomposition of the velocity and temperature fields
\begin{align}
u_i (\vec{x}) 	&= \sum_{\vec{k} \in \wavevecset{B}} \fourier{u}_i(\vec{k}) e^{i \vec{k} \cdot \vec{x}},\\
\theta(\vec{x}) 	&= \sum_{\vec{k} \in \wavevecset{B}} \fourier{\theta}(\vec{k}) e^{i \vec{k} \cdot \vec{x}},
\end{align}
and denoting $\vec{X}(\vec{k})= {^t(\fourier{u}_1(\vec{k}),\fourier{u}_2(\vec{k}),\fourier{u}_3(\vec{k}),\fourier{\theta}(\vec{k}))}$ the Fourier coefficients for wave vector $\vec{k}$, they can be expressed as a linear combination of the normal modes:
\begin{align}
\vec{X}(\vec{k}) &= A_0(\vec{k})\vec{Z}_0(\vec{k})+A_-(\vec{k})\vec{Z}_-(\vec{k})+A_+(\vec{k})\vec{Z}_+(\vec{k}),\\
& = \vec{M}(\vec{k})
\underbrace{
\begin{pmatrix}
A_-(\vec{k})\\
A_0(\vec{k})\\
A_+(\vec{k})
\end{pmatrix}}_{\vec{A}(\vec{k})}.\label{slowvareq}
\end{align}
Introducing the projectors $\vec{P}_0(\vec{k})=\vec{Z}_0(\vec{k}) \vec{Z}_0(\vec{k})^\dagger$ and $\vec{P}_W(\vec{k})=\vec{Z}_-(\vec{k})\vec{Z}_-(\vec{k})^\dagger+\vec{Z}_+(\vec{k})\vec{Z}_+(\vec{k})^\dagger$, we have of course $\vec{P}_0(\vec{k}) \vec{X}(\vec{k})=A_0(\vec{k})\vec{Z}_0(\vec{k})$ and $\vec{P}_W(\vec{k}) \vec{X}(\vec{k})=A_-(\vec{k})\vec{Z}_-(\vec{k})+A_+(\vec{k})\vec{Z}_+(\vec{k})$, and the two projectors are mutually orthogonal, so that $\vec{P}_0(\vec{k}) \oplus \vec{P}_W(\vec{k}) = I_4$.
Denoting $\mathscr{F}: (u_1,u_2,u_3,\theta) \mapsto (\fourier{u}_1,\fourier{u}_2,\fourier{u}_3,\fourier{\theta})$, with $\fourier{x}: \vec{k} \in \wavevecset{B} \mapsto \fourier{x}(\vec{k}) \in \mathbb{C}$, the Fourier transform, and $\fourier{\mathscr{P}}_0: (\fourier{u}_1,\fourier{u}_2,\fourier{u}_3,\fourier{\theta}) \mapsto (\vec{k} \mapsto \vec{P}_0(\vec{k}) \vec{X}(\vec{k}))$, $\fourier{\mathscr{P}}_W: (\fourier{u}_1,\fourier{u}_2,\fourier{u}_3,\fourier{\theta}) \mapsto (\vec{k} \mapsto \vec{P}_W(\vec{k}) \vec{X}(\vec{k}))$, the operators $\mathscr{P}_0 = \mathscr{F}^{-1} \circ \fourier{\mathscr{P}}_0 \circ \mathscr{F}$ and $\mathscr{P}_W=\mathscr{F}^{-1} \circ \fourier{\mathscr{P}}_W \circ \mathscr{F}$ are mutually orthogonal projectors acting on the set of dynamical fields:
\begin{align}
\mathscr{P}_0^2=\mathscr{P}_0 \ \ &,\ \ 
\mathscr{P}_W^2=\mathscr{P}_W,\\
\mathscr{P}_0 \mathscr{P}_W = \mathscr{P}_W \mathscr{P}_0 = 0 \ \ &,\ \ 
\mathscr{P}_0 \oplus \mathscr{P}_W = Id.
\end{align}
Hence, we may write $\vec{u}=\vec{u}_0+\vec{u}_W$ and $\theta=\theta_0+\theta_W$, with $(u_0^1,u_0^2,u_0^3,\theta_0)=\mathscr{P}_0 (u^1,u^2,u^3,\theta)$, and $(u_W^1,u_W^2,u_W^3,\theta_W)=\mathscr{P}_W (u^1,u^2,u^3,\theta)$.
It is easily checked that the projection on the slow modes defined above has vanishing vertical velocity: $u_0^3=0$, and is divergence free: $\nabla \cdot \vec{u}_0 = \nabla_\perp \cdot \vec{u}_0=0$. 
It follows from the above that the total energy 
\begin{align}
E_T&= \frac 1 2 \int (\vec{u}^2+\theta^2)
= \frac 1 2 \sum_{\vec{k} \in \wavevecset{B}} \vec{X}(\vec{k})^\dagger \vec{X}(\vec{k})
=\frac 1 2 \sum_{\vec{k} \in \wavevecset{B}} \vec{A}(\vec{k})^\dagger \vec{A}(\vec{k})
\intertext{breaks up in two pieces: $E_T=E_0+E_W$, where $E_0$ is the energy in the slow modes:}
E_0 &= \frac 1 2 \int (\vec{u}_0^2+\theta_0^2)=\frac 1 2 \sum_{\vec{k} \in \wavevecset{B}} \abs{A_0(\vec{k})}^2 = \frac 1 2 \sum_{\vec{k} \in \wavevecset{B}} E_0(\vec{k}) ,
\intertext{and $E_W$ the energy in the wave modes:}
E_W &= \frac 1 2 \int (\vec{u}_W^2+\theta_W^2)=\frac 1 2 \sum_{\vec{k} \in \wavevecset{B}} \lbrack \abs{A_-(\vec{k})}^2 + \abs{A_+(\vec{k})}^2 \rbrack = \frac 1 2 \sum_{\vec{k} \in \wavevecset{B}} E_W(\vec{k}) 
\ . \end{align}
Note that due to the orthogonality property of the projections, the cross terms in physical space cancel out: $\int \vec{u}_0 \cdot \vec{u}_W + \int \theta_0 \theta_W = 0$. However, these terms do not vanish individually. In other words,  the vector $(\mathbf{u}_0, \theta_0)$ is indeed orthogonal to the vector $(\mathbf{u}_w, \theta_w)$, but this does not imply that the vector $\mathbf{u}_0$ is orthogonal to $\mathbf{u}_w$, where orthogonality refers to the canonical scalar product in each of these two Hilbert spaces.
A consequence is that care should be taken when trying to mix decomposition of the energy in terms of kinetic/potential energy and in terms of vortical/wave modes. Here, we shall avoid altogether distinguishing between kinetic and potential energy.

\subsection{Normal mode decomposition, balance and potential vorticity}\label{balancedvorticalsection}

As can be checked explicitly~\citep[e.g.]{Herbert2014c}, in the linear framework, the slow modes satisfy balance relations: in the presence of rotation, they are in both geostrophic and hydrostatic balance. 
By contrast, for purely stratified flows, the slow modes are not in hydrostatic balance, except for purely vertical wave vectors, i.e. for horizontally homogeneous modes (the VSHF modes, see \S~\ref{vshfsection}).
Of course, nonlinear effects break these relations, and a flow initialized in the image of the projector $\mathscr{P}_0$ (i.e. containing no wave modes initially) will eventually lose balance and radiate inertia-gravity waves. 
If one is interested in maintaining the balance relations, as much as possible, without generating such oscillations, the initial state should contain a wave component designed to compensate the nonlinear effects; such \emph{nonlinear initialization} procedures have been designed in numerical weather prediction~\citep{Baer1977,Leith1980}. 
Alternatively, balanced models, where the dynamics of the waves is slaved to that of the slow modes, can be written to remain on a \emph{slow manifold} in phase space~\citep{Warn1995,Vanneste2013}. 
For the sake of simplicity, we shall be content here with projecting the data on the vector space spanned by the slow modes (which can be seen geometrically as a tangent subspace for the slow manifold, or as a zeroth order approximation to the slow manifold) and its orthogonal subspace, spanned by the wave modes, as defined above. 
This means that although the normal modes introduced above are always well defined, care should be taken when interpreting the projections in terms of balance relations.

The normal modes are also related to potential vorticity. 
An argument due in particular to~\cite{LSmith2002} relies on the fact that in the linear dynamics, $\Pi_1$ is a point-wise invariant: $\partial_t \Pi_1=0$. 
It follows by applying the Fourier transform in both space and time, that $\sigma \fourier{\Pi}_1(\vec{k},\sigma)=0$, where $\sigma$ is the time frequency. 
Hence, wave modes, for which $\sigma(\vec{k}) \neq 0$ must have $\fourier{\Pi}_1(\vec{k},\sigma)=0$, while \emph{vortical modes} (i.e., by definition, those for which $\fourier{\Pi}_1(\vec{k},\sigma) \neq 0$) must have zero frequency.
In fact, it is easily shown that in general, the identity $\fourier{\Pi}_1(\vec{k},t)=-ik\sigma(\vec{k})A_0(\vec{k},t)$ holds, and therefore~\citep{Bartello1995,Herbert2014c}:
\begin{align}
\Gamma_2 &= \sum_{\vec{k} \in \wavevecset{B}} k^2 \sigma(\vec{k})^2 \abs{A_0(\vec{k})}^2. 
\end{align}
It follows that, even when considering the full (nonlinear) dynamics, waves do not carry any (linear) potential vorticity. 
Similarly, we know that in the presence of rotation, all the slow modes contribute to (linear) potential vorticity. In the purely stratified case, however, the slow component of the VSHF modes ($k_\perp=0$) contributes neither to (linear) potential vorticity, nor, therefore, to (quadratic) potential enstrophy $\Gamma_2$. 
All the other slow modes still carry (linear) potential vorticity.

The fact that the connection between slow modes and vortical modes (i.e. modes which contribute to linear potential vorticity) does not depend on whether we are considering the linear or the full nonlinear dynamics, while the connection between slow modes and balanced modes does, is due to the \emph{kinematic} nature of the connection between linear potential vorticity and the dynamical fields.
 By contrast, the notion of balance is \emph{dynamical} in the sense that it involves pressure, which is determined by the equations of motion. 
 When including the nonlinear term in the dynamics, another component of pressure appears, which breaks the balance relations.

\section{Partition of energy between slow and wave modes} \label{S:TEMP}
\subsection{Time evolution of global quantities}\label{globalsection}

\begin{figure*}
\centering
\includegraphics[width=\linewidth]{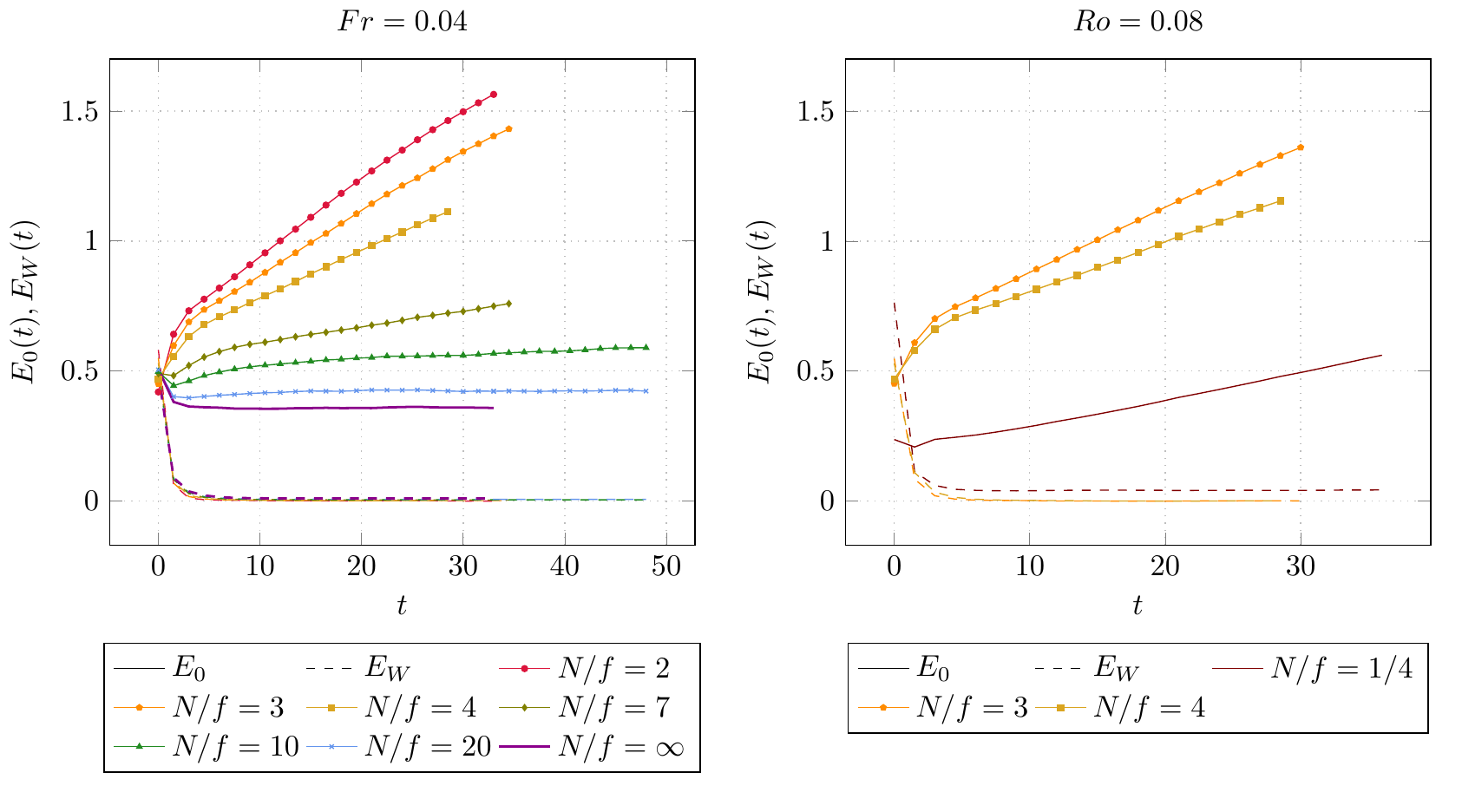}
\caption{Time evolution of the wave (dashed lines) and vortical (solid lines) parts of the total energy for the series of runs at constant Froude number (left) and at constant Rossby number (right) with various $N/f$.
Runs within the no-resonance zone have been excluded for clarity.
Note that the curves corresponding to the decay of the wave energy almost collapse onto a single curve: see Fig.~\ref{globalwavesfig} for a figure in logarithmic coordinates where they can be distinguished.
} \label{globalfig} \end{figure*}
\begin{figure*}
\centering
\includegraphics[width=\linewidth]{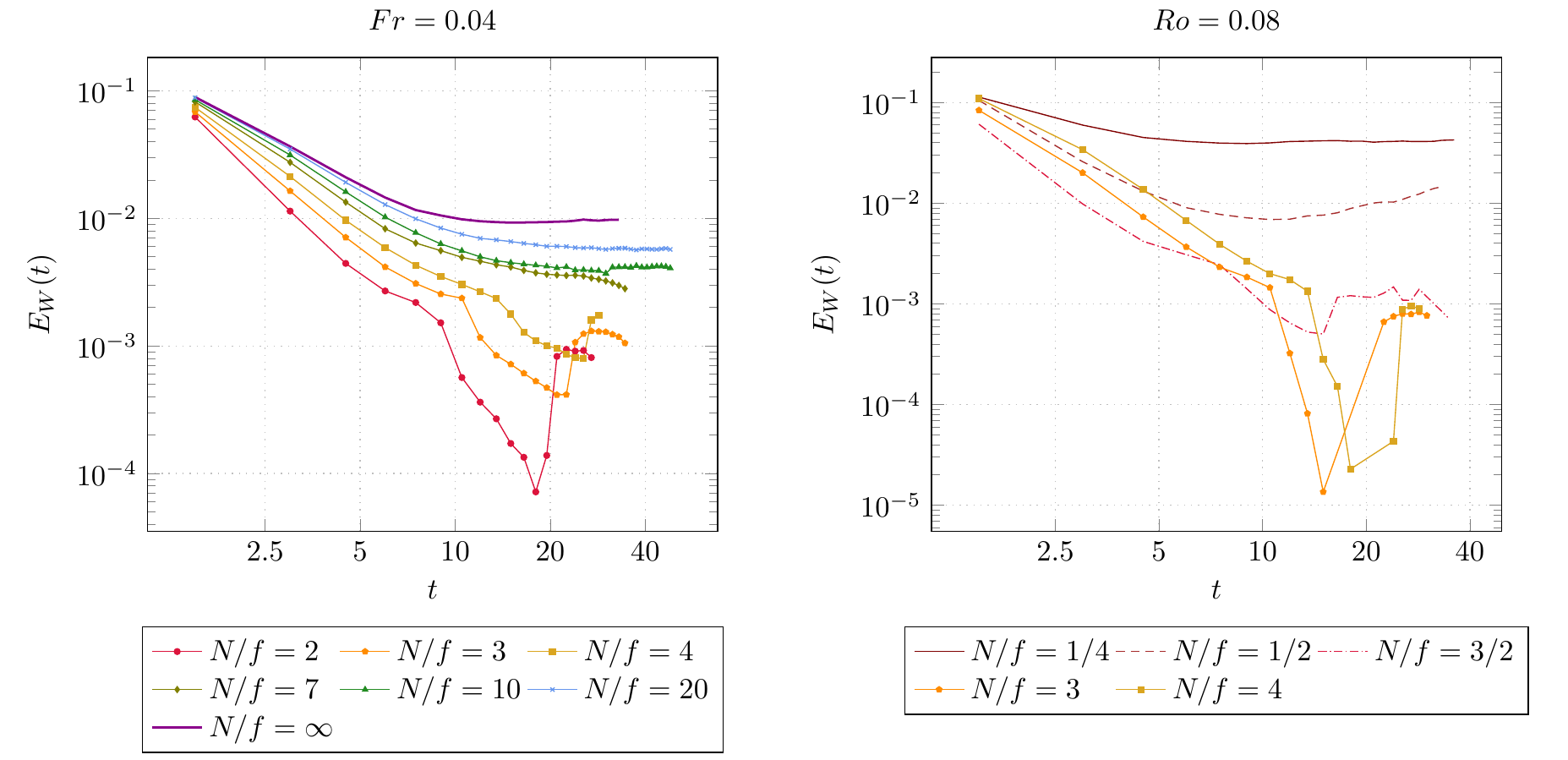}
\caption{Time evolution of the wave component of the total energy in logarithmic coordinates, for the series of runs at constant Froude number (left) and at constant Rossby number (right) with various $N/f$.
}\label{globalwavesfig} \end{figure*}

The time evolution of the two energy components, wave and vortical, is shown in Fig.~\ref{globalfig} (see also Fig. \ref{globalwavesfig}) for the series of runs at constant Froude number (left) and at constant Rossby number (right), excluding the runs for which the three-wave resonances are suppressed.
Although the initial conditions all have about the same energy in each component, we see that in all the runs, the wave energy decays very fast and the slow modes dominate, even for the runs that do not undergo a strong inverse cascade (in the purely stratified case, there is no inverse cascade at all). 
This is also true for the runs in the series at constant Rossby number (Fig.~\ref{globalfig}, right) and in the no-resonance zone (not shown). 
After an initial phase, we observe in the runs for which $N/f$ is small enough ($N/f \leq 10$) a linear growth of the energy in the slow modes, characteristic of an inverse cascade. 
\begin{figure*}
\centering
\includegraphics[height=5.25cm]{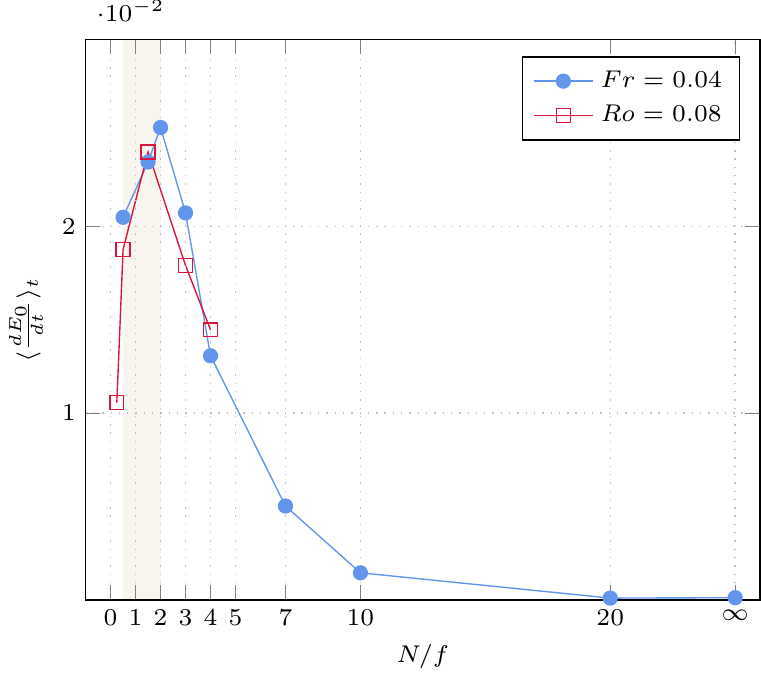}
\hfill
\includegraphics[height=5cm]{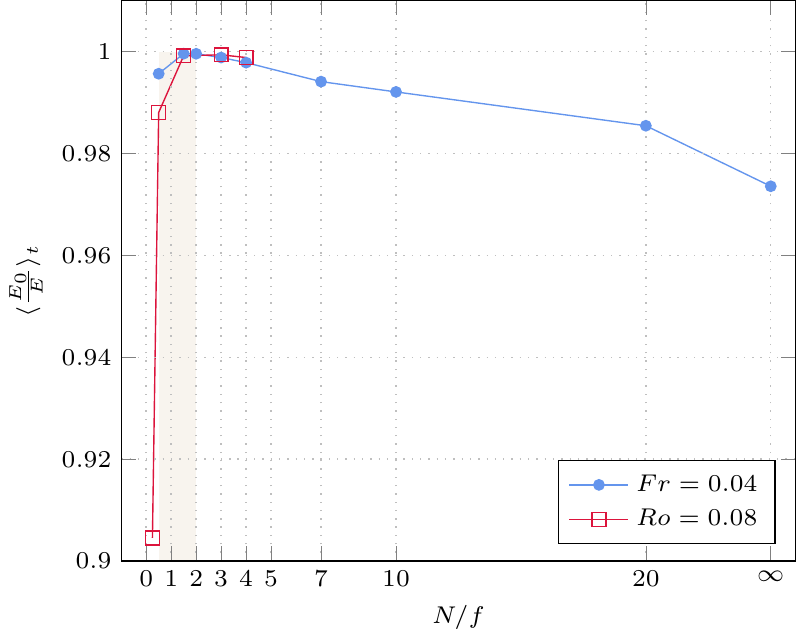}
\caption{Left:~Time-averaged growth rate of the energy in the slow modes as a function of $N/f$, for the two series of runs, at constant Froude number (circles) and at constant Rossby number (squares), respectively. 
Right:~ Time averaged fraction of energy in the slow modes as a function of $N/f$, for the two series of runs, at constant Froude number (circles) and at constant Rossby number (squares), respectively. 
The no-resonance zone is the shaded area.}
\label{globalratesfig}
\end{figure*}
For $N/f \geq 2$, the growth rate appears to be a decreasing function of $N/f$ and this is confirmed by plotting the time average of the time derivative of the energy in the slow modes (Fig.~\ref{globalratesfig}, left). 
This figure also confirms that the growth rate vanishes at $N/f=20$ or in the purely stratified case (see also \citet{Marino2014}).
There is also a linear growth of the energy in the slow modes for the runs in the no-resonance zone in the series at constant Froude and for all the runs in the series at constant Rossby. 
For both series of runs, the growth rate appears to reach a maximum located inside the no-resonance zone (Fig.~\ref{globalratesfig}, left). 
In addition, the growth rate seems to depend primarily on $N/f$ rather than the Froude and Rossby numbers. 
These conclusions are consistent with the analysis of the time evolution of the kinetic energy obtained earlier~\citep{Marino2013b}. 
Figure~\ref{globalratesfig} (right) summarizes how the fraction of energy in the slow modes, averaged in time (over all the snapshots following the peak of enstrophy), evolves as a function of $N/f$. 
Surprisingly, the case where this fraction attains its minimum value, although it is still about 0.9, is not the purely stratified case, but the rotation dominated case $N/f=1/4$, which also has the highest $\mathcal{R}_B$).
Since Fig.~\ref{globalfig} does not allow to distinguish the different curves for the wave energy at various $N/f$ ratios, we show this quantity in logarithmic scale in Fig.~\ref{globalwavesfig}. 
For values of $N/f$ above 7 (this is also true for the case $N/f=1/4$), we see a decay phase followed by a regime where the level of energy in the waves stays constant. 
For the other $N/f$ values, the wave energy drops suddenly at the end of the decay phase, then increases back to more or less its value before the drop, and seems to stay constant afterwards.
Although the time resolution is not sufficient for precise measurements, the decay seems to follow roughly a power-law, and it is steeper for smaller $N/f$ ratios (except for the $N/f=1/4$ and $N/f=1/2$ cases, at substantially higher $\mathcal{R}_B$).

\subsection{Spectral analysis}\label{specsection}

\begin{figure*}
\centering
\includegraphics[width=\linewidth]{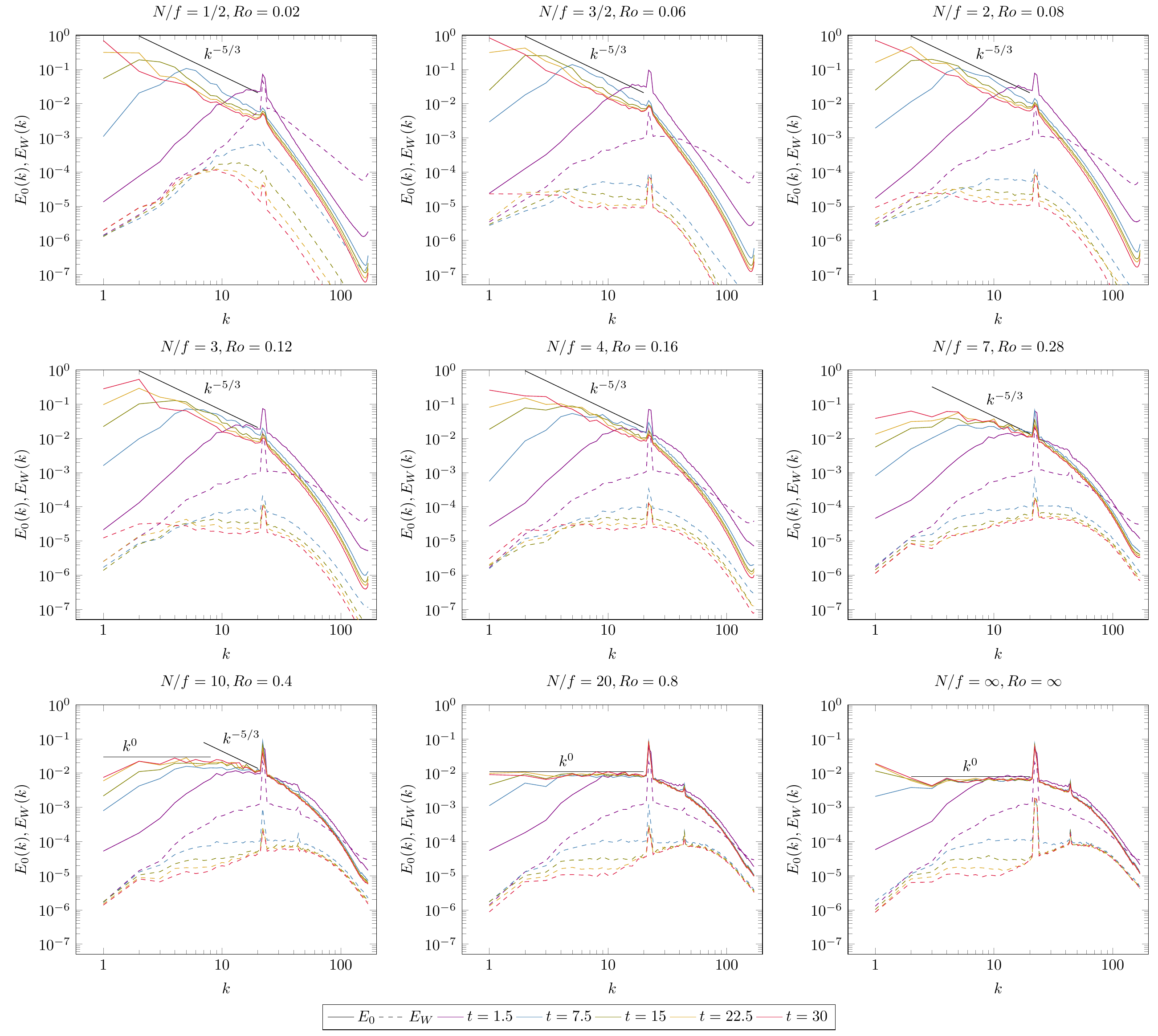}
\caption{Isotropic energy spectra of the wave (solid lines) and slow mode (dashed lines) for the series of runs with constant Froude number ($\Fr=0.04$) and varying $N/f$.
The times shown are indicated in the legend at the bottom, going from purple to blue, green, yellow and red in increasing order.
Two different spectral laws are indicated as well for comparison.
 }\label{spectraFrfig} \end{figure*}
We show in Fig.~\ref{spectraFrfig} the isotropic spectra of energy in the wave and slow modes, denoted respectively $E_W(k)$ and $E_0(k)$, for the series of runs at constant Froude number. 
The isotropic spectra are defined, as usual, as integrals over spherical shells in Fourier space: 
\begin{align}
E_0(k) &= \int_{\|\vec{k}\|=k} \abs{A_0(\vec{k})}^2 d\vec{k},& E_W(k) &= \int_{\|\vec{k}\|=k} \lbrack \abs{A_+(\vec{k})}^2 + \abs{A_-(\vec{k})}^2 \rbrack d\vec{k}.
\end{align}
For all the runs, at early times ($t=1.5$, before the peak of enstrophy), the slow mode component dominates at large scales while the wave component dominates at small scales. 
Note that the wavenumber at which the two spectra cross increases with $N/f$.
 When the turbulence develops, the wave energy spectrum decays rapidly, including at small scales, and the slow mode spectrum dominates at all scales. However, the wave energy at small scales increases when $N/f$ increases.
The other salient feature of the isotropic energy spectra is the growth of the slow mode energy at scales larger than the forcing scale. 
In most of the runs, up to about $N/f=7$, this growth takes the form of an inverse cascade, with a spectral slope that is consistent with $k^{-5/3}$. 
However, the inverse cascade of the slow modes weakens as $N/f$ increases, as already found from the analysis of global quantities. 
Already, at $N/f=7$, the spectrum at the final time of the simulation has a narrow $k^{-5/3}$ range, and at $N/f=10$, the energy in the slow modes seems to saturate at large scale with a flat spectrum $k^0$. 
Note that we cannot rule out completely the possibility that the $-5/3$ range would extend further towards large scales with a larger integration time.
However, it should be noted that the growth rate of the energy in the slow modes for these values of $N/f$ is very small (see Fig.~\ref{globalratesfig}) and that the slow mode energy spectrum have not evolved much between $t=22.5$ and $t=30$.
This indicates that even if the $-5/3$ range was to extend towards larger scales, it would require a very long integration time.
Besides, the growth does not seem to be self-similar, since the spectral slope at scales larger than the $-5/3$ range becomes shallower (almost flat), unlike traditional inverse cascade phenomena.
The flat spectrum for slow mode energy at large scales is the only notable feature of the $N/f=20$ run, and it is also present in the purely stratified run, although in that case there is an accumulation of energy in the $k=1$ modes. 
Such a flat spectrum was already observed for kinetic energy in~\citet{LSmith2002} for $N/f=10$ (see also~\citet{Kimura2012} and~\citet{Marino2013b,Marino2014}).

\begin{figure*}
\centering
\includegraphics[height=5cm]{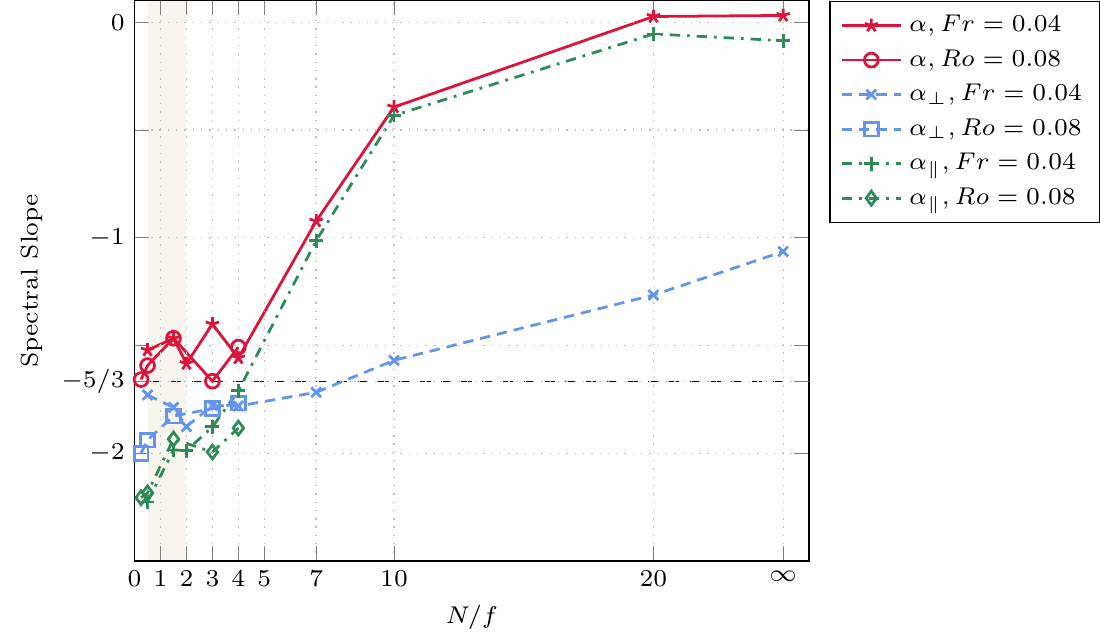}
\caption{Spectral indices for the isotropic (solid lines), perpendicular (dashed lines) and parallel (dash-dotted lines) vortical energy spectra at scales larger than the forcing, as a function of $N/f$, for the two series of runs: constant Froude number (star, $\times$ and $+$ symbols), and constant Rossby number (circle, square and diamond symbols).
The no-resonance zone is the shaded area. 
}\label{specindicesfig} \end{figure*}

The transition between these two scenarii is further described in Fig.~\ref{specindicesfig}, which shows the spectral slope $\alpha$ of the isotropic energy spectrum of the slow modes at large scale, defined by $E_0(k) \sim k^\alpha$ for $k \leq k_F$, and computed numerically by a least-square fit, as a function of $N/f$. 
This figure shows that the transition appears to be continuous, although the specific values of the spectral slopes lack in accuracy since the inertial range is rather short (remember that $k_F\in {[22,23]}$), and since they may depend on the details of the least-square fitting procedure: choice of snapshot, excluding the $k=1$ modes to get rid of the formation of a possible condensate at the largest scale,...
In particular, no time averaging has been performed here. 
Higher resolutions would also provide improvements in accuracy. 
Another caveat is that, as noted above, the intermediate values (e.g. $N/f=7$ or 10) may move closer to $-5/3$ with a larger integration time.
 
\begin{figure*}
\centering
\includegraphics[width=\linewidth]{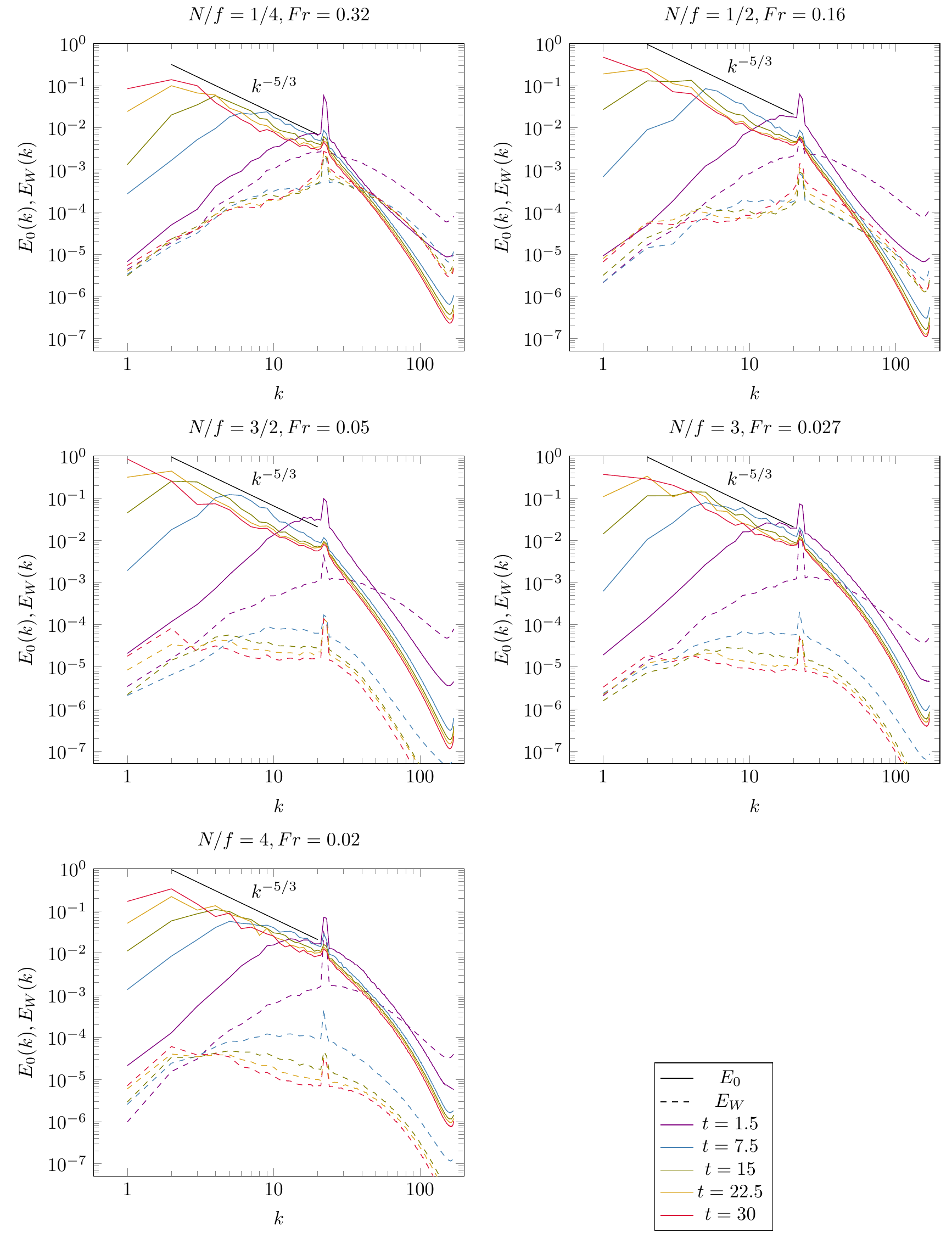}
\caption{Isotropic spectra of the wave and slow mode components of energy for the series of runs with constant Rossby number ($\Ro=0.08$) and varying $N/f$.
The symbols are the same as in Fig. \ref{spectraFrfig}.
 } \label{spectraRofig} \end{figure*}
The inverse cascade of the slow mode energy spectrum is also seen in the series of runs at constant Rossby number (Fig.~\ref{spectraRofig}).
Interestingly, for the runs with small $N/f$ ratios ($N/f=1/4$ or $1/2$), i.e. the runs with the highest Froude numbers, the waves dominate at small scales even at late times, when these scales seem to have attained stationarity. 
This is probably a consequence of the higher buoyancy Reynolds number, rather than the Burger number (see the discussion in \S~\ref{rbsection}).

\begin{figure*}
\centering
\includegraphics[width=\linewidth]{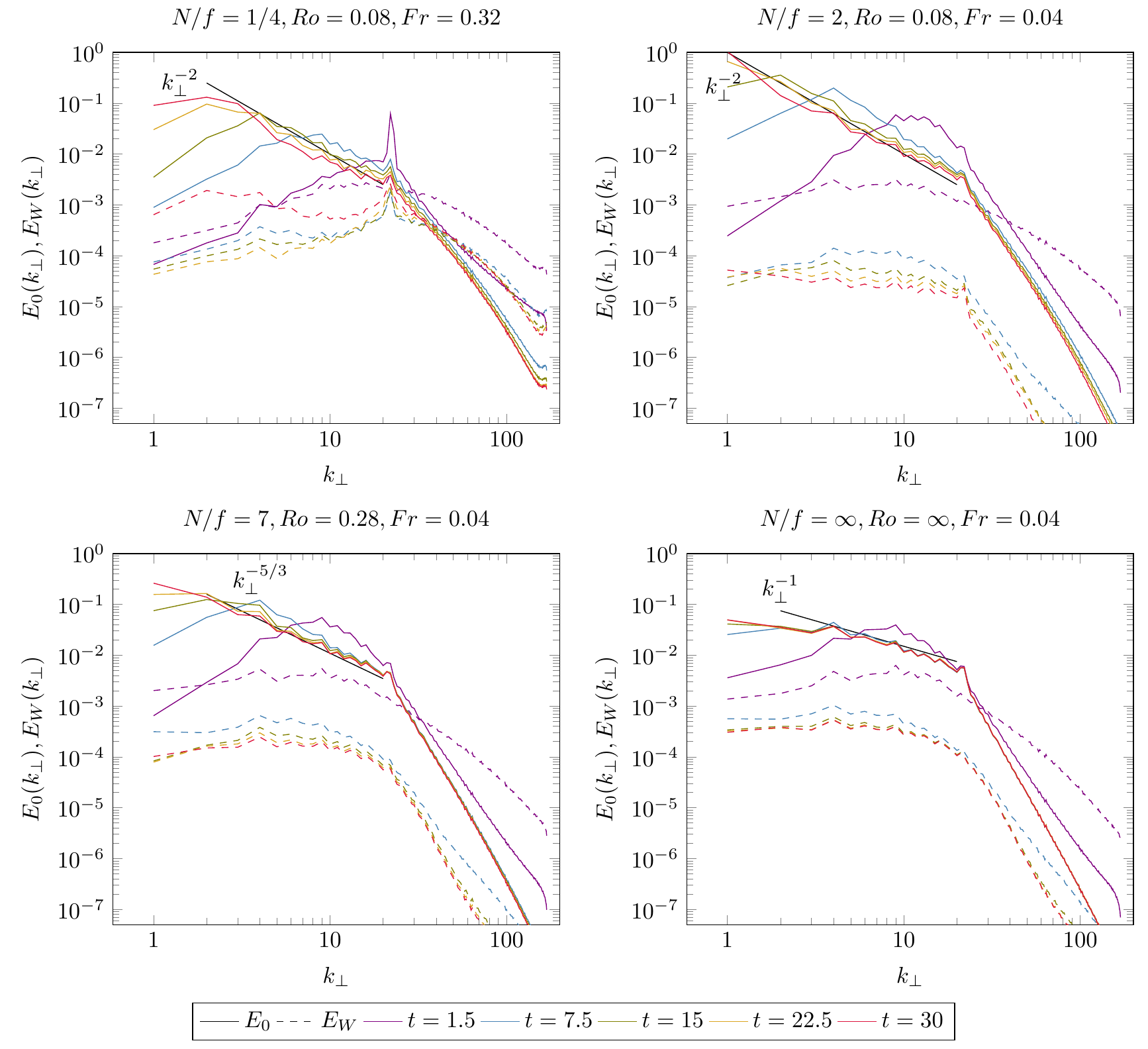}
\caption{Perpendicular spectra of the wave and slow mode components of energy for selected values of the parameters. Scaling laws are represented only for reference. 
Line encoding and colors are similar to the two preceding figures.
}\label{specperpfig} \end{figure*}
In addition to the isotropic spectrum, obtained by summing over spherical shells in Fourier space, it is always relevant for anisotropic flows to compute the so-called \emph{reduced spectra}, where the summation is done either on cylinders (fixed $k_\perp$) or planes (fixed $k_\parallel$) in Fourier space~\citep[see e.g.][]{Godeferd2003,SagautCambonBook,Mininni2011c,Mininni2012,Marino2014}. 
Introducing the axisymmetric spectra $e_0(k_\perp,k_\parallel),e_W(k_\perp,k_\parallel)$ defined by
\begin{align}
e_0(k_\perp,k_\parallel) &= \int_0^{2\pi} \abs{A_0(k_\perp \cos \phi \vec{e}_x + k_\perp \sin \phi \vec{e}_y + k_\parallel \vec{e}_z)}^2 k_\perp d\phi,\\
e_W(k_\perp,k_\parallel) &= \int_0^{2\pi} \lbrack \abs{A_+(k_\perp \cos \phi \vec{e}_x + k_\perp \sin \phi \vec{e}_y + k_\parallel \vec{e}_z)}^2 + \abs{A_-(k_\perp \cos \phi \vec{e}_x + k_\perp \sin \phi \vec{e}_y + k_\parallel \vec{e}_z)}^2 \rbrack k_\perp d\phi,
\end{align}
the perpendicular and parallel spectra are defined as follows for each type of mode, $s=0,W$:
\begin{align}
E_s(k_\perp) &= \int_0^{k_{\text{max}}} e_s(k_\perp,k_\parallel)dk_\parallel,& E_s(k_\parallel) &= \int_0^{k_{\text{max}}} e_s(k_\perp,k_\parallel)dk_\perp.
\end{align}
Figure~\ref{specperpfig} shows the perpendicular spectrum of energy in the wave ($E_W(k_\perp)$) and slow modes ($E_0(k_\perp)$); again, except for the runs with the highest Froude number ($Fr=0.32$ and $Fr=0.16$, not shown), and therefore, buoyancy Reynolds number, the slow modes dominate at all horizontal scales after the peak of enstrophy. 
In all cases, the perpendicular spectrum of the slow modes undergoes a growth of energy in horizontal scales larger than the forcing scale (note, however, that since the forcing acts on a spherical shell in Fourier space, there is some injection of energy at all the horizontal scales such that $0 \leq k_\perp \leq k_F$). 
As for the isotropic spectrum, this growth appears to be self-similar for $N/f$ ratios smaller than 10. As $N/f$ increases, the perpendicular spectrum saturates at large horizontal scales; see the purely stratified case in Fig.~\ref{specperpfig} for instance. 
The spectral slope in this range varies depending on the non-dimensional parameters, and it is difficult to provide precise values here. 
Nevertheless, it appears that the perpendicular spectrum becomes shallower as $N/f$ increases. 
 The spectral index $\alpha_\perp$, defined by $E_0(k_\perp) \sim k_\perp^{\alpha_\perp}$ for $k_\perp \leq k_F$, and computed numerically by a least-square fit, is represented as a function of $N/f$ in Fig.~\ref{specindicesfig}. 
 The caveats mentioned above regarding accuracy of this spectral index still apply here.
For instance a slightly steeper spectral slope might have been obtained in the purely stratified case if we had chosen a range of wave numbers closer to the forcing wavenumber: \citet{Marino2014} claim that the kinetic energy (here we are showing the slow mode energy) has a $-5/3$ slope in this range.
The perpendicular wave spectrum differs markedly in the rotation-dominated, weakly stratified case and in the strongly stratified cases. 
In the former case, more energy remains at small horizontal scales, while in the latter case it falls off sharply at scales smaller than the forcing scales, and it is shallow or flat at scales larger than the forcing scales.

\begin{figure*}
\centering
\includegraphics[width=\linewidth]{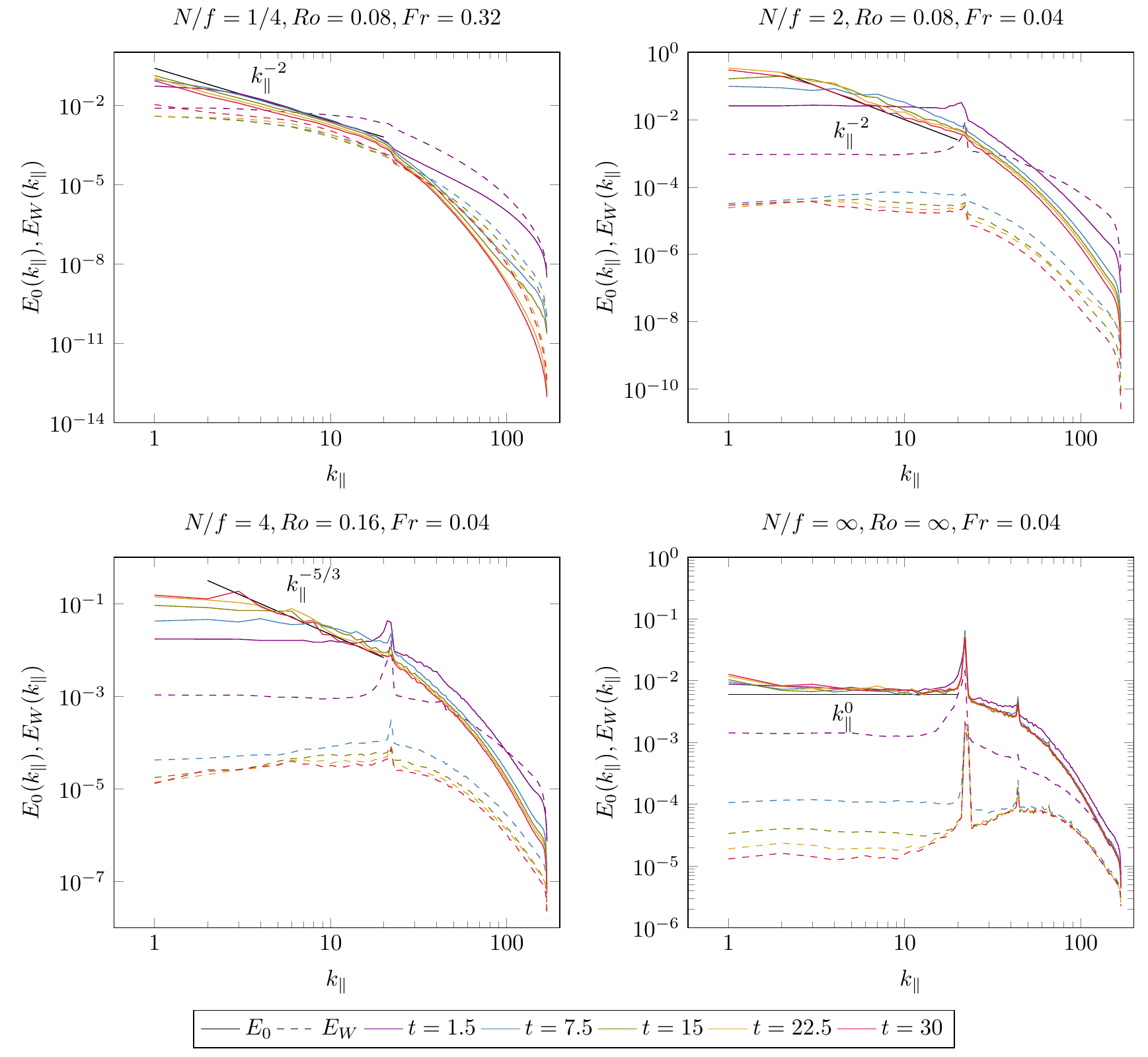}
\caption{Parallel spectra of the wave and slow mode components of energy for selected values of the parameters. 
Scaling laws are represented only for reference. 
Line encoding and colors are similar to the three preceding figures.
}\label{specparafig} \end{figure*}
Similarly, the parallel spectra $E_W(k_\parallel), E_0(k_\parallel)$ of energy in the wave and slow modes, respectively, are shown in Fig.~\ref{specparafig}. 
Again, once the turbulence sets in, the slow modes dominate at all vertical scales except in the two cases with high buoyancy Reynolds numbers, where the waves dominate at small-scales (for the case $N/f=1/4$, the waves dominate roughly at all the vertical scales smaller than the forcing scale $k_F$, although the energy in these scales decay extremely fast). 
Overall, the energy in the small scales increases as we move from the rotation dominated case (where the spectrum is very steep) to the purely stratified case (note the different vertical scales on the panels of Fig.~\ref{specparafig}). 
This statement also holds for the wave component of the energy: while the waves have most of their energy at large-scales for rotation dominated cases or cases where rotation and stratification are comparable (i.e. relatively low $N/f$ ratios), they become mostly small-scale when stratification prevails.
The buildup of slow mode energy at large vertical scales occurs differently from the perpendicular spectra: already at very early times, the parallel spectrum of the slow modes is more or less flat at scales larger than the forcing scale. 
Besides, for the lowest $N/f$ ratio ($1/4$), there is relatively little change in the slow mode parallel spectrum at large scales.
 As for the perpendicular spectra, there does not seem to be a universal scaling law in this range: the spectral slope $\alpha_\parallel$ (defined by $E_0(k_\parallel) \sim k_\parallel^{\alpha_\parallel}$ for $k_\parallel \leq k_F$) varies between about $-2$ and $-5/3$ until $N/f=4$ (see Fig.~\ref{specindicesfig}), then the spectrum becomes much shallower at large scales, and it is almost flat for the purely stratified case and for the case with $N/f=20$.

\subsection{Role of the VSHF modes}\label{vshfsection}

Several studies of stratified flows have underlined the role of horizontally homogenous modes:~\cite{Herring1989} and~\cite{Staquet1998} have reported a tendency towards layering, while~\citet[see also \cite{Godeferd1994} for an EDQNM perspective]{LSmith2002} have shown that at very long time, energy accumulates in the horizontally homogeneous modes (defined by $k_\perp=0$), with a peak at some value of $k_\parallel$ which defines the number of horizontal layers in the vertical, hence the name \emph{vertically sheared horizontal flows} (VSHF). 
 Our numerical simulations are stopped earlier in order to avoid the accumulation of energy in the gravest mode of the system that is inherent to all inverse cascade phenomena (see e.g.~\cite{Boffetta2012} for a recent review in 2D); this provides an incentive for looking at these modes in the direct numerical simulations carried out here.

\begin{figure*}
\centering
\includegraphics[height=5cm]{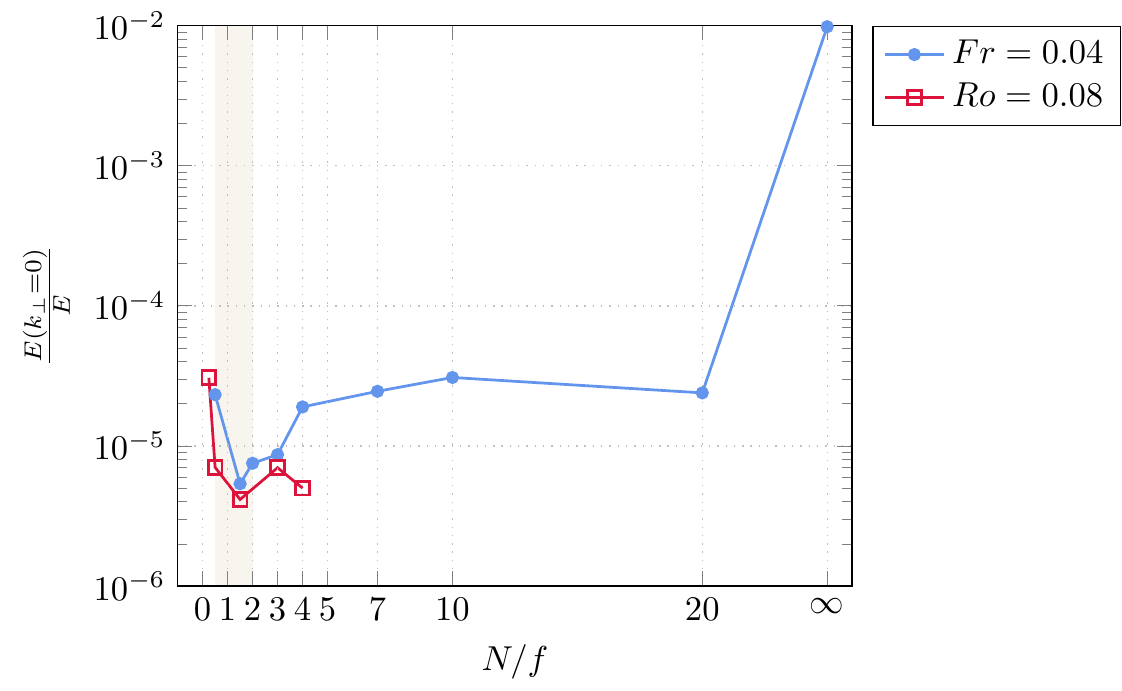}
\caption{
Fraction of energy in the vertically sheared horizontal modes ($k_\perp=0$ modes) as a function of $N/f$, for the two series of runs. The no-resonance zone is the shaded area.
}\label{globalvshffig}
\end{figure*}
By definition, the VSHF modes are horizontally homogeneous, but may depend on the vertical coordinate.
In general, they have an inertial wave component (whose frequency is the Coriolis frequency $f$) corresponding to an oscillation of the horizontal components of velocity, with vanishing vertical velocity, and a slow mode component that is proportional to temperature fluctuations. 
Hence, their wave component is purely kinetic and their slow mode component is purely potential. 
When rotation vanishes, the dispersion relation also vanishes and there is no oscillatory component anymore.

The ratio of energy in these VSHF modes is shown in  Fig.~\ref{globalvshffig}, as a function of $N/f$ for the two series of runs.
It is striking that it remains of the same order of magnitude (and relatively small; it is in fact negligible compared to the waves for instance, see Fig.~\ref{globalwavesfig}) for all finite values of $N/f$; it is only when we move to the strictly non-rotating case that the energy in the VSHF modes jumps by about two orders of magnitude. 
Then, the energy in the $k_\perp=0$ modes becomes comparable to the energy in waves with any other wave vector.
Note that, although we do not have enough runs at high buoyancy Reynolds number for a systematic study, it does not seem to be a relevant parameter as far as this issue is concerned since the two runs with higher buoyancy Reynolds have a similar level of energy in the VSHF modes as the other runs.

\subsection{Energy fluxes}\label{transfersection}

Let us now turn to the fluxes of energy in the different modes. We first define the transfer functions $T_0(\vec{k})$ and $T_W(\vec{k})$:
\begin{align}
T_0(\vec{k}) &= \Re \left\lbrack \vec{X}_0(\vec{k})^\dagger \sum_{\vec{p}+\vec{q}=\vec{k}} i k^j \fourier{u}_j(\vec{p}) \vec{X}(\vec{q}) \right\rbrack,\\
T_W(\vec{k}) &= \Re \left\lbrack \vec{X}_W(\vec{k})^\dagger \sum_{\vec{p}+\vec{q}=\vec{k}} i k^j \fourier{u}_j(\vec{p}) \vec{X}(\vec{q}) \right\rbrack,
\end{align}
where we have introduced the projections $\vec{X}_0(\vec{k})=\vec{P}_0(\vec{k}) \vec{X}(\vec{k})$ and $\vec{X}_W(\vec{k})=\vec{P}_W(\vec{k}) \vec{X}(\vec{k})$.
The evolution of the energy in the slow and wave modes at wave vector $\vec{k}$ is simply governed by
\begin{align}
(\partial_t+2\nu k^2) E_0(\vec{k}) &= T_0(\vec{k}),\\
(\partial_t+2\nu k^2) E_W(\vec{k}) &= T_W(\vec{k}).
\end{align}
As usual, we can define the isotropic, perpendicular and parallel transfer functions $T_s(k)$, $T_s(k_\perp)$ and $T_s(k_\parallel)$ (with $s=0,W$) by integrating over a spherical shell, a plane or a cylinder in Fourier space. 
\begin{figure*}
\centering
\includegraphics[width=\linewidth]{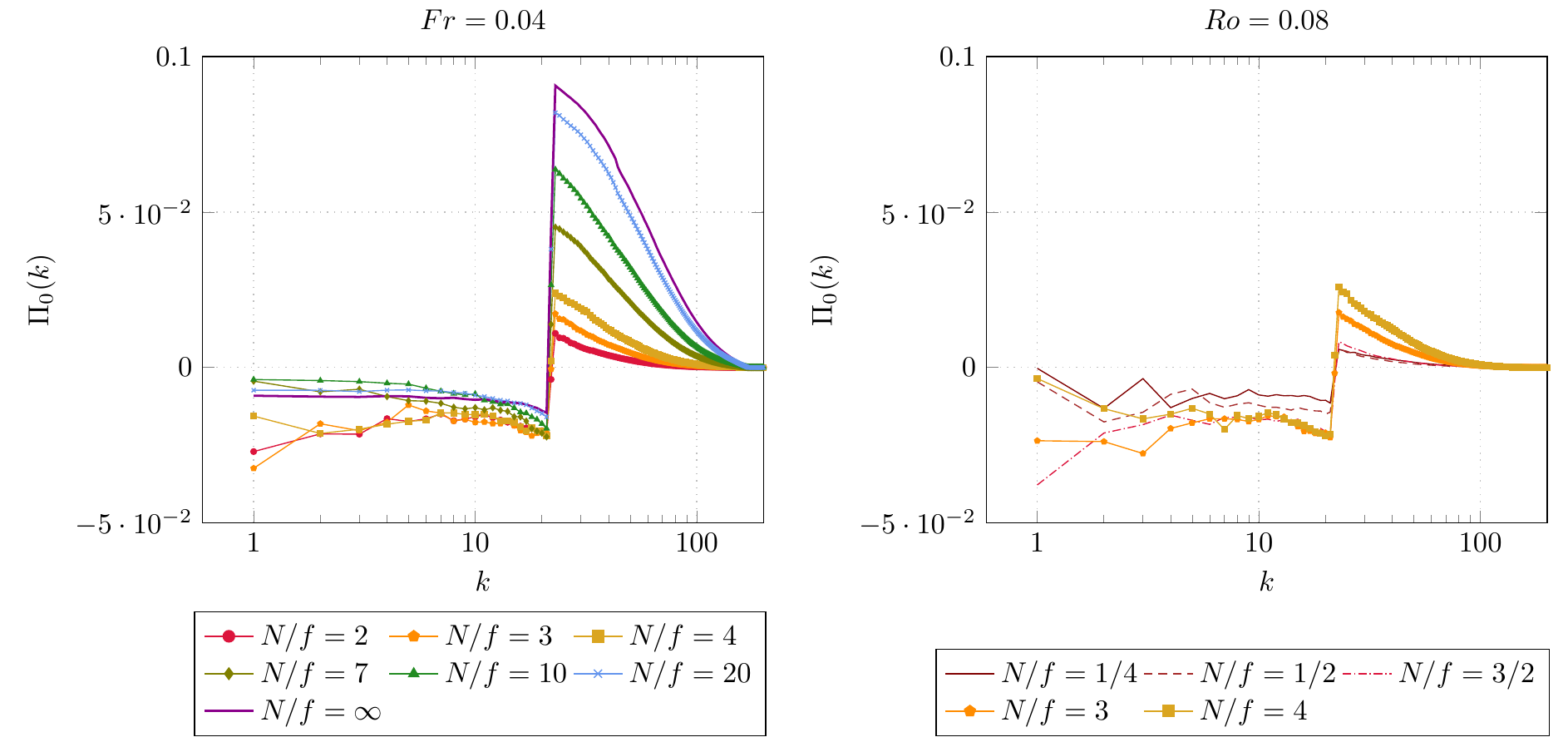}
\includegraphics[width=\linewidth]{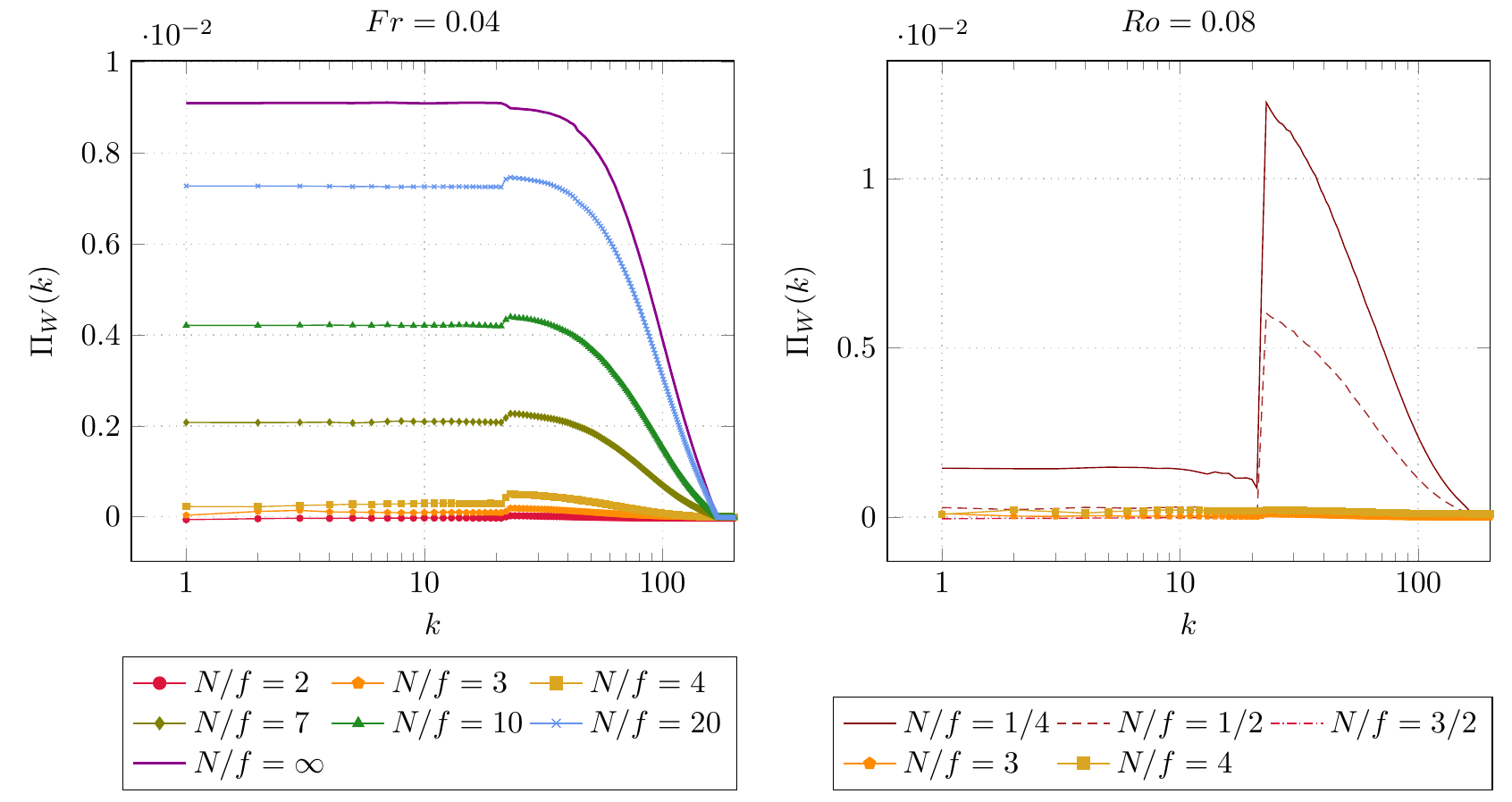}
\caption{Isotropic fluxes of energy for the slow modes (top) and for the wave modes (bottom) for the series of runs at constant Froude number (left) and at constant Rossby number (right) with various N/f. 
Note the different scale (they differ by a factor 10) for $\Pi_0$ and $\Pi_W$.
}\label{isofluxfig}
\end{figure*}
Then we introduce the fluxes:
\begin{align}
\Pi_0(k) &= \int_{k}^{k_{\text{max}}} T_0(k)dk,\\
\Pi_W(k) &= \int_{k}^{k_{\text{max}}} T_W(k)dk,
\end{align}
and similarly for the perpendicular and parallel fluxes $\Pi_s(k_\perp)$ and $\Pi_s(k_\parallel)$.

The isotropic energy fluxes for the slow modes and for the waves are shown in Fig.~\ref{isofluxfig} for the latest time ($t=30$).
We see that for all $N/f$ ratios, the energy fluxes for the slow modes are negative, and more or less independent of $k$, at scales larger than the forcing. 
Furthermore, the magnitude of this negative flux seems to be larger for relatively small values of $N/f$ (the runs with $N/f=2,3,4$ have about the same magnitude), while it is reduced when stratification prevails.
On the contrary, the fluxes are positive at scales smaller than the forcing.
In this range, the positive flux is an increasing function of $N/f$
On the other hand,  the energy fluxes for the wave modes do not change sign at the forcing scale: they are always positive.
The absence of a range with constant small-scale fluxes can be attributed to the small size of the $k>k_F$ wave number range, a problem remedied when using higher resolutions and/or lower forcing wavenumber \citep{Pouquet2013, Marino2015}, or a parametrization scheme \citep{Pouquet2013b}.
Note that unlike the traditional (total) energy fluxes, we need not have $\Pi_s(0)=0$, because $E_0$ and $E_W$ are not independently conserved (only their sum is).
This means that care should be taken when interpreting fluxes, since the value of the flux for one given mode (slow or wave) at $k$ is not only the amount of energy transferred from wave numbers smaller than $k$ to wave numbers larger than $k$, but also includes the contributions from the other mode.
\begin{figure*}
\centering
\includegraphics[height=5cm]{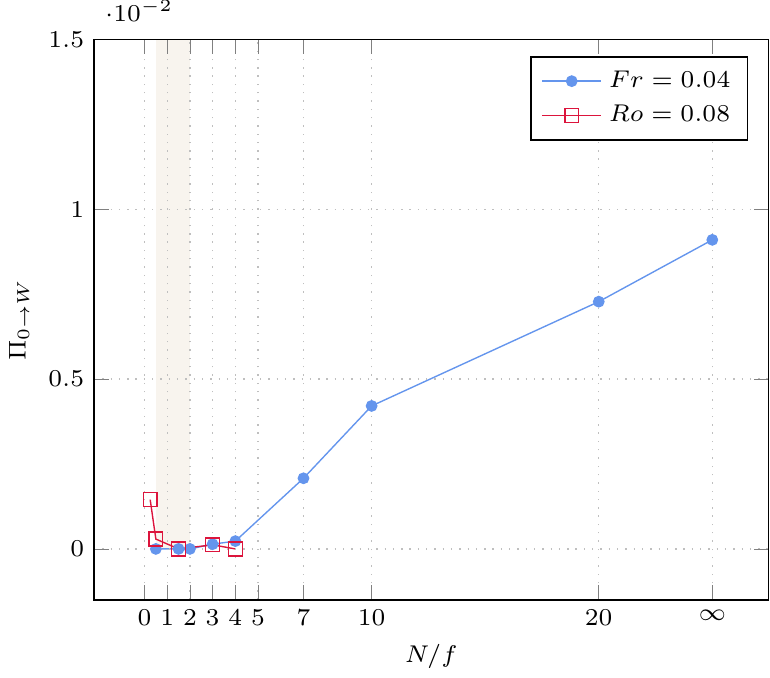}
\caption{Total energy flux from the slow modes to the wave modes as a function of $N/f$.
}\label{flux0Wfig}
\end{figure*}
The total flux of energy from the slow modes to the waves $\Pi_{0 \to W}$, which coincides with $\Pi_W(0)$, or equivalently $-\Pi_0(0)$ (up to numerical errors), is shown as a function of $N/f$ in Fig.~\ref{flux0Wfig}. 
It is small when $N/f$ is small (up to $N/f=4$) and then increases almost linearly with $N/f$.
For all values of $N/f$ larger than unity, the energy flux for the wave modes (Fig.~\ref{isofluxfig}) is never significantly larger than $\Pi_{0 \to W}$, which means that the transfer of wave energy is dominated by the leakage from slow modes.
However, note that for the two cases $N/f=1/4$ and $N/f=1/2$, the flux at scales smaller than the forcing scale is much larger than $\Pi_{0 \to W}$, meaning that in those cases, there is a significant transfer of energy towards the small-scales from the waves themselves.
This interpretation is consistent with the fact that in the $N/f=1/4$ case the energy in the wave modes is significantly larger than in any other case, and these two cases are those for which the waves dominate at small scales (see Fig.~\ref{spectraRofig}), and for which the buoyancy Reynolds number is significantly larger.
Similarly, the negative flux at scales larger than the forcing scales for the slow modes is on the order of $\Pi_{0 \to W}$ when $N/f \geq 7$. This means that the negative flux alone does not allow to claim with certainty that we have an inverse cascade of slow modes.
For instance, slow energy in the stratified case has a negative flux while we have seen above that it lacks some characteristics of the inverse cascade.
What is more meaningful is the fact that in the range of scales between the forcing scale and the plateau of the flux at the largest scales, the flux decreases with $k$ and goes below the level of $\Pi_{0 \to W}$, all the more so that $N/f$ is small, consistently with the above analysis.
For the runs with $N/f \leq 4$, the interpretation of the negative flux of slow energy is much easier: the magnitude of the more or less constant flux over scales larger than the forcing scales is much larger than $\Pi_{0 \to W}$, which means that it can safely be attributed to an inverse cascade of slow mode energy.

\begin{figure*}
\centering
\includegraphics[width=\linewidth]{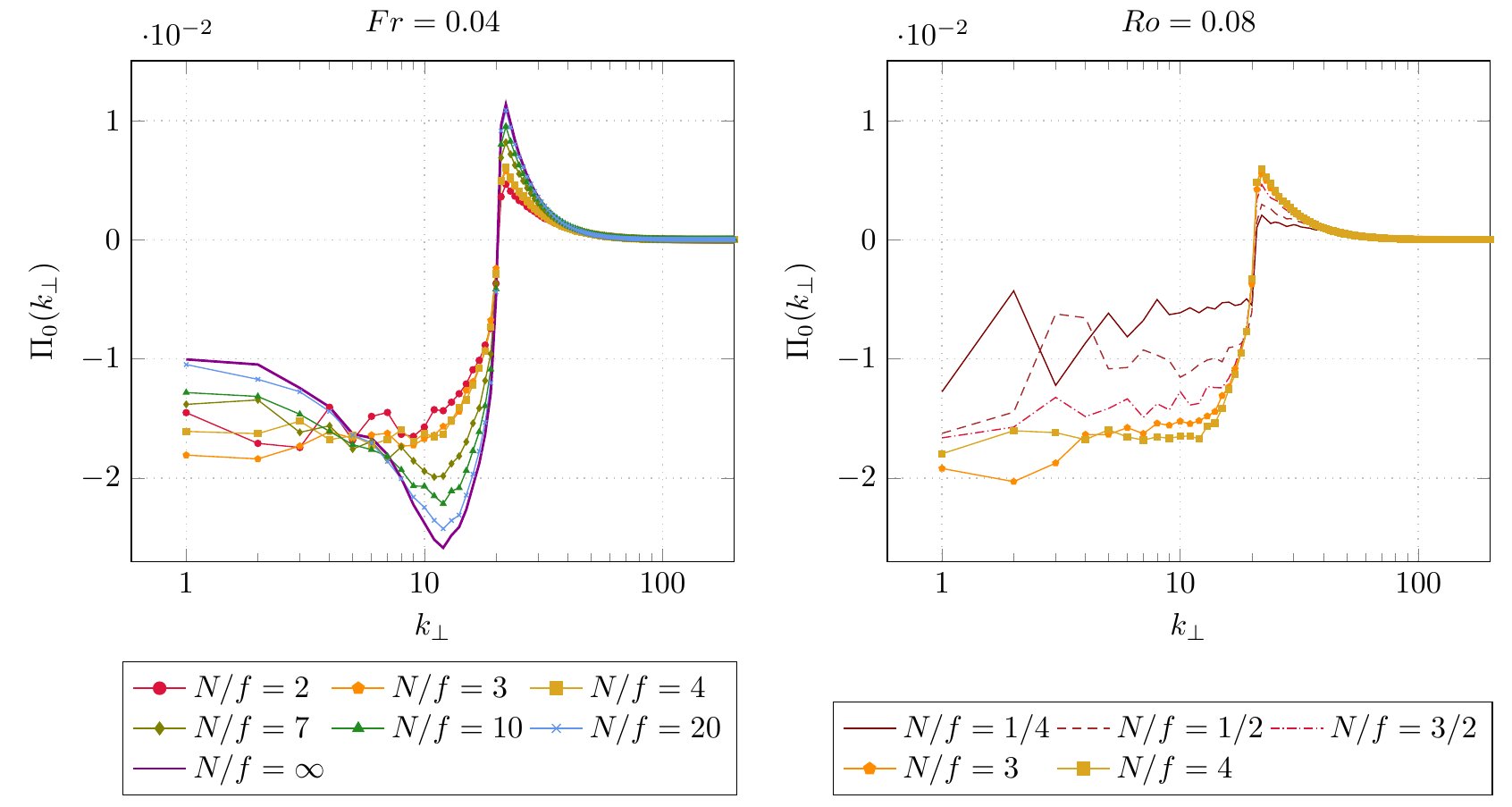}
\includegraphics[width=\linewidth]{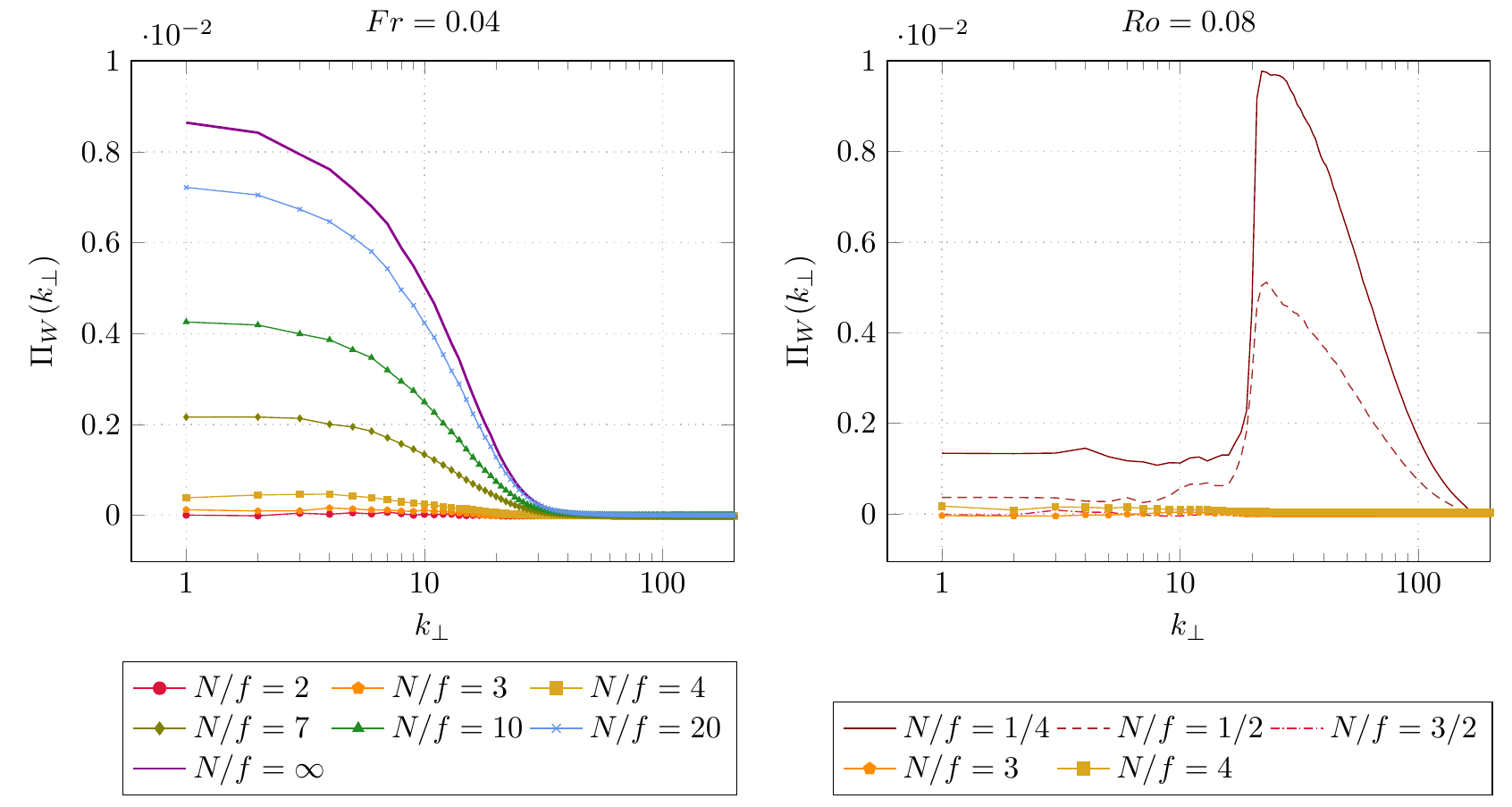}
\caption{Perpendicular fluxes of energy for the slow modes (top) and for the wave modes (bottom) for the series of runs at constant Froude number (left) and at constant Rossby number (right) with various N/f.}\label{perpfluxfig}
\end{figure*}
In Fig.~\ref{perpfluxfig}, we show the perpendicular fluxes of energy in the slow modes and in the waves, $\Pi_0(k_\perp)$ and $\Pi_W(k_\perp)$.
For the slow modes, the fluxes are similar to the isotropic fluxes, with a notable difference: the negative fluxes at scales larger than the forcing scale are enhanced.
In particular, now all the runs have a wide range of scales (larger than the forcing scale) for which the flux has larger magnitude than $\Pi_{0 \to W}$. 
That means that regardless of $N/f$, slow mode energy tends to be transferred towards large horizontal scales.
This is known to be true even for purely stratified flows or flows strongly dominated by rotation, which tend to form uniform layers~\citep{Riley2000}.
As a matter of fact, runs dominated by stratification have a (negative) peak in the slow mode energy flux whose magnitude increases with $N/f$.
The fluxes indicate, however, that for stratification dominated runs (e.g. $N/f=20$ or $\infty$), this transfer of slow energy is arrested at some horizontal scales (like kinetic energy; see~\cite{Marino2014}), while it continues all the way to the largest scales when stratification is weaker (e.g. $N/f=2,3,4$ or even 7 and 10).
On the contrary, the energy flux in the wave modes are always smaller than $\Pi_{0 \to W}$, and contrary to the isotropic fluxes, they do not have any range of constant value. 
This indicates that the transfer of energy from the slow modes to the waves probably occurs at large horizontal scales (we cannot in principle draw definitive conclusions since there might be cancellations between energy transfer from slow to wave modes and transfer across scales of wave modes, but it does not seem very plausible that they are of different signs).
Note that the forcing scale is difficult to identify from this figure alone.
It should be emphasized that since we are forcing in a spherical shell in Fourier space, energy is actually injected in the full range $k_\perp \in [0,k_F]$ (the same goes for $k_\parallel$ of course).
Hence a caveat in the interpretation of anisotropic fluxes is that the forcing scale is not so well defined.
However, in almost all cases except the above mentioned, the forcing scale can still be identified unambiguously in $\Pi_s(k_\perp)$ and $\Pi_s(k_\parallel)$, which gives us confidence that the forcing is not contaminating the results too much.
Finally, the perpendicular fluxes of wave energy for the two cases $N/f=1/4$ and $N/f=1/2$ are very similar to the isotropic fluxes, which means that in those cases, the near-inertial waves tend to transfer their energy towards small horizontal scales.

\begin{figure*}
\centering
\includegraphics[width=\linewidth]{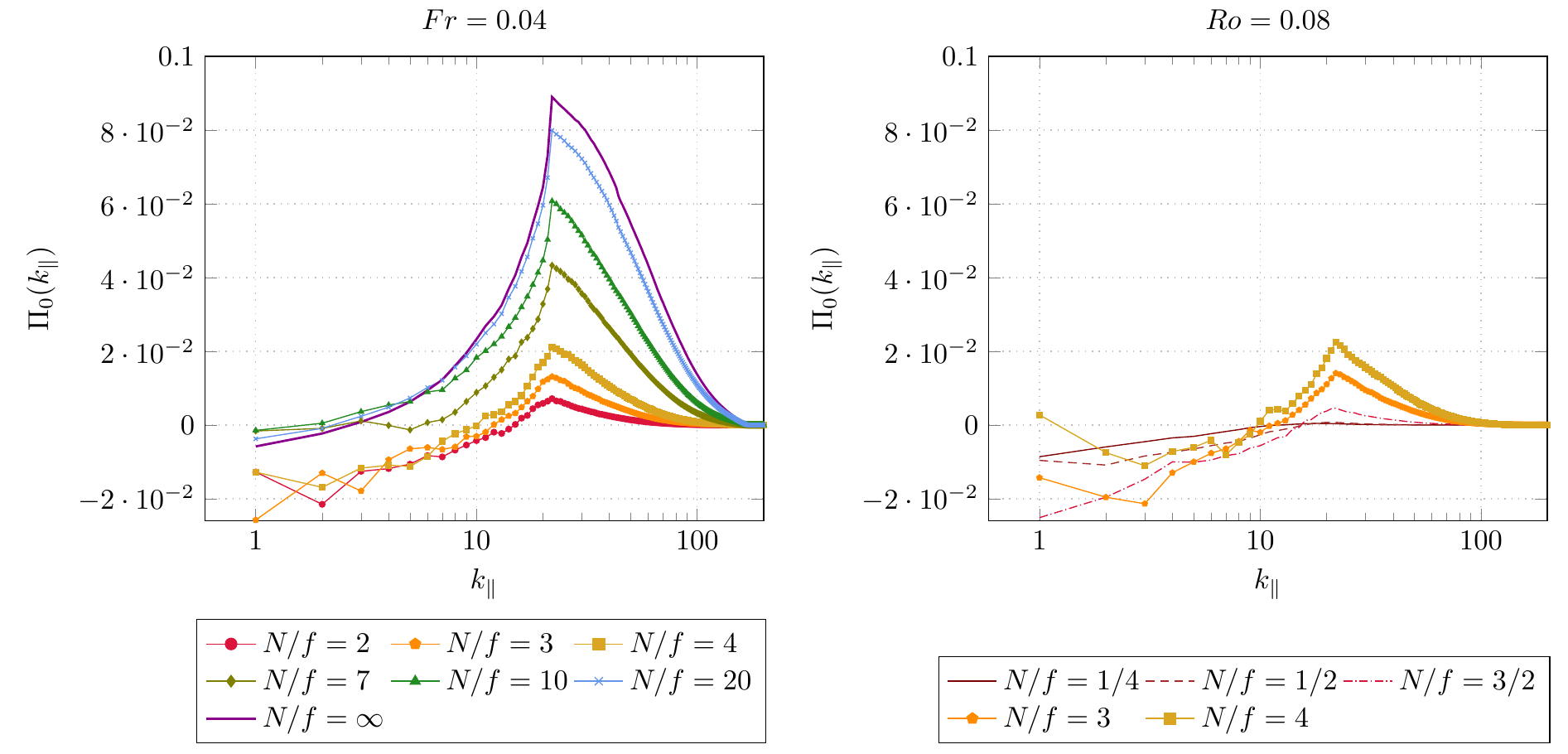}
\includegraphics[width=\linewidth]{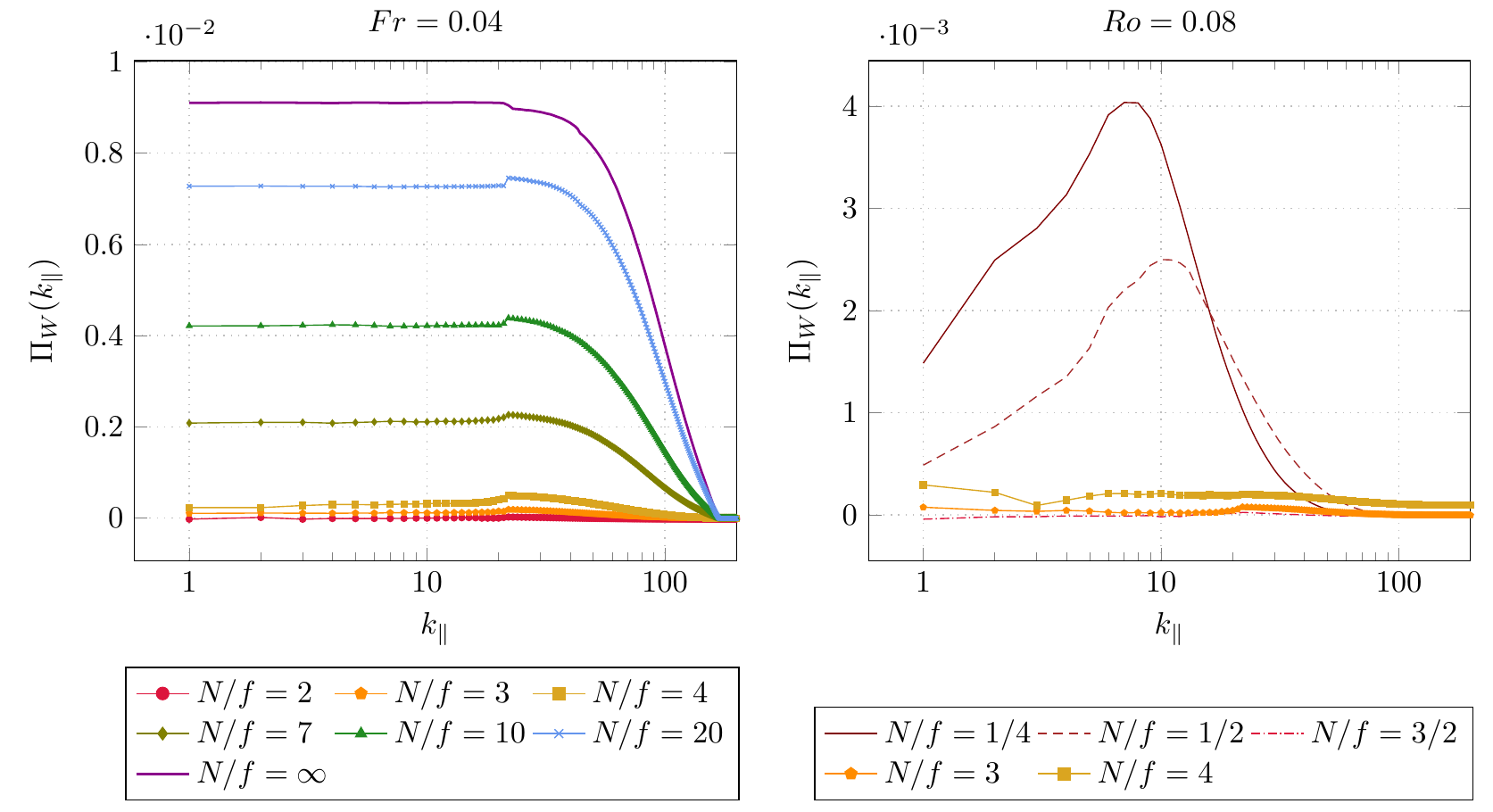}
\caption{Parallel fluxes of energy for the slow modes (top) and for the wave modes (bottom) for the series of runs at constant Froude number (left) and at constant Rossby number (right) with various N/f.}\label{parafluxfig}
\end{figure*}
Now, we turn to the examination of the parallel flux of energy in the slow modes and in the wave modes (Fig.~\ref{parafluxfig}).
The most notable results are that runs with $N/f \geq 7$ have positive fluxes of slow energy across the whole range of vertical scales, corresponding to the maintenance of horizontal layers, whose thickness decreases with $N/f$.
This process is less clear when stratification is weaker, since the parallel flux then remains negative at the very large scales.
For all the runs with $N/f \geq 2$, the parallel flux of wave energy is identical to the isotropic flux, indicating that vertical transfers dominate, although interpretation is still polluted by the exchange with slow modes.
When $N/f=1/4$ or $N/f=1/2$, it is interesting to note that there is a peak at vertical scales larger than the forcing scale (this is another case where the isotropic forcing scale does not appear clearly), although its magnitude is relatively small.
This peak does not correspond to any identifiable feature in the isotropic fluxes.

\section{Discussion}\label{discusssection}

\subsection{Summary of the results}\label{sumconclusec}

To sum up, \S~\ref{globalsection},~\ref{specsection},~\ref{vshfsection} and~\ref{transfersection} have established that in all the runs shown here:
\begin{itemize}
\item The total energy is dominated by the slow mode component, even when stratification is strong, although the fraction of energy in the waves increases when $N/f \geq 2$ increases. 
This happens in spite of the fact that the forcing has roughly equal projections on the slow and wave modes.
The increase in wave energy with $N/f$ is consistent with the observed total flux of energy from the slow modes to the wave modes, which also increases with $N/f$.
\item The slow mode component dominates at all scales, except in the two runs with higher buoyancy Reynolds number (those with $N/f=1/4$ and $1/2$), where they only dominate at large scale. 
This also holds when distinguishing horizontal and vertical scales.
\item When $N/f$ is small enough (roughly $N/f \leq 7$, although the specific value may depend on the details of the setup, and, in particular, the buoyancy Reynolds number and the strength of the waves due to rotation and/or stratification), the slow mode component has all the characteristics of an inverse cascade of energy: linear growth of their global energy, self-similar growth of their energy at scales larger than the forcing scales, and negative and constant fluxes at scales larger than the forcing scale.
\item When $N/f$ is higher, this inverse cascade of energy is arrested. 
When stratification prevails, slow modes dominate the total energy, and there is some transfer of these modes towards the large scales, but it does not take the form of an inverse cascade: it saturates with a flat spectrum at large isotropic and vertical scales, and there is no significant negative flux of energy for those scales. 
The spectrum is not flat at large horizontal scales, consistent with the presence of significant negative flux of slow mode energy, but it is still much shallower than in the presence of rotation.
\item As long as rotation is present, the fraction of energy in horizontally homogeneous modes (VSHF modes, satisfying $k_\perp=0$) does not vary much with $N/f$, and this fraction is always much lower than the energy in the waves, which is itself much lower than the energy in the slow modes. 
On the contrary, in the purely stratified case, the VSHF modes reach a level of energy comparable with the waves.
Interestingly, this transition appears to be pretty sharp, contrary to all the other properties of the flows we have studied here (e.g. global partition of energy between waves and slow modes, growth rate of the slow mode energy and spectral indices), which seem to have a smoother transition as $N/f$ increases.
\end{itemize}

\subsection{The role of slow modes, waves and VSHF modes for the direction of the energy cascade}\label{cascadesection}

One aspect of the results outlined above (\S~\ref{sumconclusec}) which is of particular interest is that the discrepancy between the phenomenologies of rotating-stratified flows (inverse cascade of energy, when there is sufficient rotation), and purely stratified flows (direct cascade of energy) does not seem to be explained only by the different role of waves but also by the intrinsic difference in the behaviour of the slow modes in these two cases.

A framework which is quite robust for understanding direct and inverse cascades is that of the inviscid invariants of the system. 
More precisely, the existence of a second quadratic, definite positive invariant, in addition to energy, may prevent the downscale cascade of energy and lead to an inverse cascade. This is what happens in 2D turbulence~\citep{Kraichnan1980}. 
In stratified turbulence, with or without rotation, there is such a second inviscid invariant, though not quadratic in principle: potential enstrophy (see \S~\ref{pvsection}). 
Although this invariant exists regardless of whether there is rotation or not, the phenomenology is not the same. 
A subtlety here is that the invariant has constant sign (it is always positive), but is not definite: some combinations of the fields make potential vorticity vanish. 
Such combinations correspond to the wave modes. 
Hence, a natural way of thinking is to say that if waves and slow modes did not exchange energy, the wave and slow mode energy would be separately conserved, and slow modes would have invariants similar to 2D turbulence, while waves would have invariants similar to 3D Homegenous Isotropic Turbulence. 
Therefore, in that case one should expect an inverse cascade of slow mode energy and a direct cascade of wave energy.

However, in the purely stratified case, this reasoning does not hold since then, potential enstrophy is still degenerate even when we restrict it to the slow modes. 
Indeed, in the non-rotating case, only slow modes with $k_\perp\neq 0$ contribute to potential enstrophy (see \S~\ref{balancedvorticalsection}). 
This peculiarity was used in a statistical mechanics argument~\citep{Herbert2014c} in order to explain why the two cases have opposite cascade directions in spite of having the same inviscid invariants.
From this point of view, it could also be because of the presence of the \emph{vertically sheared horizontal flow} modes that stratified turbulence does not have an inverse cascade. 
A similar situation occurs in homogeneous isotropic turbulence, where the second invariant, helicity, is not sign-definite in general (in our case the sign of potential enstrophy is constant but it can still vanish for some modes). 
When one decimates the system by keeping only helical modes of a given sign, helicity becomes sign-definite and an inverse cascade results, as observed numerically~\citep{Biferale2012} and in agreement with statistical mechanics arguments~\citep{Zhu2014,Herbert2014a}. 
But when one starts reintroducing in the dynamics a fraction of the modes which destroy sign-definiteness of the invariant, the inverse cascade collapses very soon~\citep{Sahoo2015}, which again seems in agreement with the statistical mechanics prediction~\citep{Herbert2014a} and the behavior of purely stratified flows described above. 
An alternative constraint which leads to the reversal of the energy flux is confinement: when the aspect ratio of the domain is small, enstrophy production is inhibited at scales larger than the layer thickness, even in the presence of stratification~\citep{LSmith1996,Celani2010,Sozza2015}.
To further characterize a potential mechanism involving the VSHF modes which prevents the inverse cascade in the purely stratified case, it would be interesting to look at triadic exchanges in more details. 
This is beyond the scope of this paper, and it is left for future studies.

\subsection{Variation with buoyancy Reynolds number} \label{rbsection}

As pointed out by several authors~\citep{Brethouwer2007,Ivey2008, Bartello2013, Waite2013b}, the buoyancy Reynolds number $\mathcal{R}_B$ plays an important role in stratified turbulence, and it is likely doing so as well when rotation is  included. 
In particular, the regime of low buoyancy Reynolds number corresponds to a case where horizontal layers are viscously coupled. 
It is characterized by a vertical scale set by the viscous scale $\sqrt{\nu L_h/U}$ rather than the buoyancy scale $U/N$~\citep{Riley2003,GodoyDiana2004,Brethouwer2007,Rorai2014}, a small vertical Froude number (on the contrary, at high buoyancy Reynolds number, the vertical Froude number remains of order unity~\citep{Billant2001}) and a potential enstrophy which is dominated by its quadratic component~\citep{Waite2013b}. 
Here we have explicitly verified the latter property (see \S~\ref{pvsection}). 
Hence, the results presented here tend to show that in this regime, the system relaxes to a balanced state where the velocity field is almost horizontal (the slow modes have vanishing vertical velocity), and slow mode energy is cascaded upscale provided rotation is strong enough.

For higher $\mathcal{R}_B$, data from the two weakly stratified runs point to the fact that waves now dominate the small-scales. 
This is bound to happen when the Ozmidov scale is resolved, since then isotropy is recovered and there is no longer any reason for the velocity field to be mostly horizontal: since the vertical velocity in the normal mode decomposition framework is entirely accounted for by the wave modes, their contribution has to be larger at these scales. 
Note that we have seen in \S~\ref{pvsection} that in this regime, the nonlinear components of potential vorticity are also stronger (see Fig.~\ref{gammaratiofig}).
In this high $\mathcal{R}_B$ regime, numerical simulations have shown that both an inverse and a direct cascade of energy were possible~\citep{Pouquet2013, Marino2015}. 
From the point of view of the inviscid invariants of the system, this is a puzzling result since inverse cascades usually arise when a second invariant prevents the direct cascade (see \S~\ref{cascadesection}). 
The key to this phenomenon presumably lies in the fact that the energy breaks up into a wave component (not constrained by potential enstrophy conservation, as explained above) and a vortical component (constrained by potential enstrophy conservation), although it could be expected that in this regime, more energy is transferred from the slow modes to the waves than in the case investigated in the present paper. 
However, as soon as one resolves the inverse cascade, the Reynolds number, and therefore $\mathcal{R}_B$ are limited to moderate values since the forcing wavenumber is necessarily relatively large. 
Hence, such a study can only be done at high resolution. 
We leave this analysis for future work.

\section{Conclusion} \label{S:CONCL}

We have studied the partition of energy in terms of waves and vortices, defined through the normal modes of the linearized equations, in stratified flows with or without rotation. 
We have focused on the inverse cascade regime, which appears when rotation is sufficiently strong but not when stratification dominates. 
Using numerical simulations covering a wide range of $N/f$ ratios, we have characterized this transition in terms of the waves and vortices.

Our main finding is that in all cases, the total energy is dominated by the vortices, and in most cases (i.e. when the buoyancy Reynolds number is of order one) this is true at all scales, vertical or horizontal alike. 
We have shown that in the range of parameters where the inverse cascade is present (i.e. when the ratio $N/f$ is small enough), the vortices themselves exhibit all the characteristics of the inverse cascade: linear increase of the global energy in time, self-similar growth of the energy spectrum at scales larger than the forcing scale, and negative and (approximately) constant energy fluxes in the inertial range. 
When $N/f$ becomes larger (here because rotation is decreased), these features disappear, in agreement with the weakening of the inverse cascade observed previously~\citep{Marino2013b,Marino2014}.

This is a puzzling result in that the inviscid invariants, which are traditionally reliable indicators of the direction of the energy cascade, are the same, regardless of the $N/f$ ratio, including in the purely stratified case. 
There are several elements which may contribute to the weakening, and ultimately, disappearance, of the inverse cascade of slow modes as $N/f$ increases. 
On the one hand, although vortices dominate in all cases, the total flux of energy from slow modes to waves increases as $N/f$ increases (see Fig.~\ref{flux0Wfig}), and consequently the fraction of energy in the waves also increases (see Fig.~\ref{globalwavesfig}). 
Besides, in the purely stratified case, there is an accumulation of energy in the VSHF modes, which do not contribute to quadratic potential enstrophy (see Fig.~\ref{globalvshffig}). 
Finally, as $N/f$ increases, the approximation of potential enstrophy by its quadratic part, on which the inviscid invariants argument is based, becomes less and less accurate (see Fig.~\ref{gammaratiofig}). 
Future research would be needed to fully disentangle the relative role of these three effects.

\textit{The work of C.~H. on this research was partially supported by the \emph{Advanced Study Program} and the \emph{Geophysical Turbulence Program} at NCAR.
The National Center for Atmospheric Research is sponsored by the National Science Foundation.
R.~M. was supported by the Regional Operative Program, Calabria ESF 2007/2013 and the European Community's Seventh Framework Programme FP7-PEOPLE-2010-IRSES under grant agreement n. 269297 - TURBOPLASMAS. 
A.~P. acknowledges support from the Laboratory for Atmospheric and Space Physics, University of Colorado.
D.~R. acknowledges support from the Oak Ridge Leadership Computing Facility at ORNL via the Office of Science under DOE Contract No. DE-AC05-00OR22725.
Computer time was provided by ASD/ASP (NCAR).}

 \end{document}